\documentclass[pre, twocolumn, showpacs, groupaddress, showkeys, floatfix]{revtex4-1}

\usepackage{amsmath}
\usepackage{amsfonts}
\usepackage{amssymb}
\usepackage{graphicx}
\usepackage{dcolumn}
\usepackage{bm}
\usepackage{soul, xcolor}
\usepackage{color, wasysym}
\usepackage{textcomp}
\usepackage{srcltx}
\usepackage{appendix}
\usepackage{latexsym}
\usepackage{epsfig}
\usepackage{mathrsfs}

\DeclareMathOperator{\D}{d\!}
\DeclareMathOperator{\E}{e} 
\DeclareMathOperator{\I}{i}

\newtheorem{theorem}{Theorem}
\newtheorem{definition}{Definition}
\newtheorem{lemma}{Lemma}

\newtheorem{remark}{Remark}

\begin{document} 


\allowdisplaybreaks

\makeatletter

\title{Operational solutions for the generalized Fokker-Planck and \\ generalized diffusion-wave equations} 

\author{K. G\'{o}rska}
\email{katarzyna.gorska@ifj.edu.pl}
\affiliation{Institute of Nuclear Physics, Polish Academy of Science, \\ ul. Radzikowskiego 152, PL-31342 Krak\'{o}w, Poland}


\begin{abstract}
The evolution operator method is used to solve the generalized Fokker-Planck equations and the generalized diffusion-wave equations in the (1+1) dimensional space in which $x\in\mathbb{R}$ and $t\in\mathbb{R}_+$. These equations contain either the first- or the second-time derivatives smeared by memory functions, each of which forms an integral kernel (denoted by $f(\xi, t)$, $\xi\in\mathbb{R}_+$) of suitable evolution operators. If memory functions in the Laplace space are Stieltjes functions, then $f(\xi, t)$ satisfy normalization, non-negativity, and infinite divisibility to be considered a probability density function. The evolution operators also contain exponential-like operators whose action on initial condition $p_0(x) > 0$ leads to the parent process distribution functions. This makes the results fully analogous to those obtained within the standard subordination approach. The above conclusion is satisfied by the solution of the generalized Fokker-Planck equation. In the case of the generalized diffusion-wave equation, to get this property, we should employ a special class, namely "diffusion-like" initial conditions. The key models of the operator method involve power-law memory functions. It leads to the characterization of $f(\xi, t)$ by applying one-sided stable L\'{e}vy distributions. The article also examines the properties of evolution operators in terms of evolution and self-reproduction.
\end{abstract}


\allowdisplaybreaks

\maketitle

\section{Introduction}\label{sect0}

Many real-world systems display anomalous behavior that cannot be adequately described by simple models of kinetic phenomena, such as those focused on standard diffusion or wave behavior.  Anomalies are characteristic of systems with increasingly complex structures, where interactions among constituents and inhomogeneities in the environment play a crucial role. The physical consequences can be time delayed response of the system to external stimuli. To describe such behavior, we need to go beyond evolution equations rooted in the locality principle and replace them with their time non-local generalizations. For instance, the diffusion case is described either by the non-local diffusion equation or by the non-local wave equation \cite{FMainardi96, YuLuchko14}. This non-locality is described by the so-called memory function, which is assumed to be a non-increasing function with a locally integrable singularity at $t=0$ and vanishes at $\pm$infinity. 

Typically, memory functions are represented by power-law and distributed order functions \cite{KGorska20, TSandev19, TSandev17}. When the memory function is expressed by a power-law function, the range of the exponent corresponds to sub-, normal, and super-diffusion. These types of diffusion are frequently observed in biological cells \cite{CIonescu17} and tumor growth \cite{AlineiT23}, among the other phenomena. In the distributed order memory function scenario, the system exhibits super-slow and super-fast diffusion \cite{AVChechkin02, AVChechkin03}. Such behavior is observed in the Sinai model \cite{YGSinai82}, in mechanical DNA unzipping \cite{YKafr06, JCWaler12}, aging environments processes \cite{MALomholt13}, annealed continuous time random walks with logarithmic waiting time distribution \cite{AGodec14}, and similar applications. The fractional derivatives are employed in the description of the wave propagation as well. For instance, the electromagnetic waves in plasma can be studied by the so-called Caputo generalized fractional derivatives \cite{NBhangale20} or the Atangana-Baleanu derivatives \cite{KBKachhia21}. The conformal derivative is used to describe the dynamical behavior of water waves \cite{AAMamun24}. Plenty of applications of the 'fractional' dynamics with different memory functions providing the anomalous diffusion and wave propagation are given by \cite{HGSun18, VUchaikin18, RHilfer00} and references therein.

The local evolution equations in classical physics can be solved by using the Green functions. For example, the solution of diffusion, wave, and telegraph (called the Cattaneo-Vernotte in heat transfer) equations obtained from this method are widely known and can be found in seminal books as, e.g., \cite{PMMorse53, JAStratton41, FWByron-v2}. Having in mind the subordination procedure \cite{KGorska20, KGorska23a, EBazhlekova19, AVChechkin21, FMainardi03, IMSokolov02}, we can also obtain the solution of their anomalous version. Through the subordination approach, the solutions to the anomalous versions of these equations involve the integration of the product of two functions: the probability density function (PDF) of the parent process and the independent PDF of the leading process. The first PDF is represented by the solution of a local (not blurred) equation. The latter PDF encodes information about the memory kernels.

Another method of solving the diffusion and wave equation involves utilizing the evolution operator formalism, similar to its application in solving the time-independent Schr\"{o}dinger equation. According to this approach the solution of the (1+1)-dim. diffusion equation, $N(x, t)$, can be presented as 
\begin{align}\label{21/05/24-4}
\begin{split}
N(x, t) & = \E^{B t \partial_x^2} q_0(x) \\
& = \int_{-\infty}^{\infty} \D y\, \frac{1}{2 \sqrt{\pi B t}} \E^{-\frac{(x-y)^2}{4 B t}} q_0(y),
\end{split}
\end{align}
where $x\in\mathbb{R}$, $B$ is a diffusion coefficient, and $q_0(x)$ is a initial condition, see \cite[Chapter III.9]{DVWidder75}. The solution of the (1+1)-dim. wave equation $W(x, t)$ reads 
\begin{align}\label{12/01/24-7}
\begin{split}
&W(x, t) = {\cos(a t \partial_{x})} p_{0}(x) + {\sin(a t \partial_{x})} v_{0}(x) \\
& = \frac{1}{2}\big[p_{0}(x + a t) + p_{0}(x - at)\big] 
+ \frac{1}{2 a} \int_{x - a t}^{x + at} \D y\, v_{0}(y)
\end{split}
\end{align}
with ${\sin(a t \partial_{x})} v_{0}(x) = \int_{0}^{t} \D\xi\, {\cos(a \xi \partial_{x})} v_{0}(x)$, see \cite[Corollary 3.14.8]{WArendt11}. A wave propagation speed is denoted as $a$ whereas the initial conditions are equal to $W(x, t=0) = p_{0}(x)$ and $\dot{W}(x, t)|_{t=0} = v_{0}(x)$. 

Up to now, the operational method approach was also used for solving the fractional Fokker-Planck \cite{EBazhlekova19, KGorska12} and the diffusion-wave equations \cite{EBazhlekova18} in which the memory functions have power-law forms. The solution of the (1+1)-dim. fractional Fokker-Planck equation, $\big({^{C\!} D}^{\alpha}_t q\big)(\alpha; x, t) = L_{F\!P}\, q(\alpha; x, t)$ where $x\in\mathbb{R}$, $t\in\mathbb{R}_{+}$, and $\alpha\in(0, 1)$, can be obtained by acting the evolution operator ${\cal E}(\alpha; t)$ on the initial condition $q_0(x)$ \cite{F0}. Formally, ${\cal E}(\alpha; t)$ is defined by the Mittag-Leffler function $E_{\alpha}(t^\alpha L_{F\!P})$ and it is represented by 
\begin{equation}\label{21/05/24-1} 
{\cal E}(\alpha; t) = \int_0^\infty \D\xi\, f(\alpha; \xi, t) \, {\cal E}(1; \xi), \qquad {\cal E}(1; \xi) = \E^{\xi L_{F\!P}},
\end{equation} 
{where} the Fokker-Planck operator {$L_{F\!P} \equiv L_{F\!P}(x, \partial_x)$ {is equal to} $B \partial_x^2 - \mu \partial_x$} with $B > 0$ being a diffusion coefficient and $\mu \geq 0$ { being a drift} coefficient. The symbol ${^C D}^{\alpha}_t$ represents the fractional derivative in the Caputo sense; see Appendix \ref{app1}. The function $f(\alpha; \xi, t)$ for $\alpha\in(0, 1)$ reads 
\begin{equation}\label{21/05/24-2} 
f(\alpha; \xi, t) = \frac{t}{\alpha \xi} \varPhi_{\alpha}(\xi, t), \quad \xi > 0 \quad \text{and} \quad t > 0,
\end{equation} 
where the two-arguments L\'{e}vy stable distribution $\varPhi_{\alpha}(\xi, t) = \mathscr{L}^{-1}[\E^{-\xi s^{\alpha}}; t]$ serves as an example of the Wright-Mainardi function \cite{EBazhlekova19, FMainardi03, FMainardi07, FMainardi10}. By the self-reproducing property, it can be expressed in terms of the one-sided L\'{e}vy stable distribution with one argument, that is, $\varPhi_{\alpha}(\xi, t) = \xi^{-1/\alpha} \varPhi_{\alpha}\big(t \xi^{-1/\alpha}\big)$. The explicit form of the one-sided L\'{e}vy stable distribution with one argument can be found in \cite{ELukacs70, KGorska21b, TSandev23book} or in Appendix \ref{app2}. 

The solution of the (1+1)-dim. diffusion-wave equation, $\big({^{C\!} D}^{2\beta}_t p\big)(2\beta; x, t) = \partial_{x}^{2}\, p(2\beta; x, t)$ where $\beta\in(1/2, 1)$, presented in the form of evolution operator ${\cal U}(2\beta; t)$ acting on the initial condition $p_{0}(x)$ is exhibited in the mathematical paper \cite{EBazhlekova18}. Obtained there evolution operator ${\cal U}(2\beta; t)$ is defined as
\begin{equation}\label{30/10/24-1}
{\cal U}(2\beta; t) = \int_{0}^{\infty} f(\beta; \xi, t) {\cal U}(2; \xi), \quad {\cal U}(2; \xi) = {\cos(\xi \partial_{x})}
\end{equation}
and is related to the subordination approach since the second initial condition, i.e. $\dot{p}(x, t)|_{t=0}$, vanishes. The auxiliary function $f(\beta; \xi, t)$ appearing in Eq. \eqref{30/10/24-1} above is the same as given by Eq. \eqref{21/05/24-2} but now for $\beta\in(1/2, 1)$.

In this paper, we will apply the evolution operator formalism to solve the generalized Fokker-Planck equation (abbreviated as GFPE) \\
\begin{equation}\label{21/05/24-3}
     \int_0^t  M(t-\xi) \partial_{\xi} q_{s\hat{M}}({x}, \xi) \D\xi = L_{F\!P}\, q_{s\hat{M}}({x}, t) 
\end{equation}
and the generalized diffusion-wave equation (abbreviated as GDWE)
\begin{equation}\label{8/01/24-1}
\int_{0}^{t} k(t-\xi) \partial_{\xi}^{2} p_{s\hat{k}}({x}, \xi) \D\xi = a^{2} \partial_{x}^{2}\, p_{s\hat{k}}({x}, t)  
\end{equation}
where ${x}\in\mathbb{R}$ and $t\in\mathbb{R}_{+}$. For the power-law memory functions both GFPE and GDWE go to their fractional versions called the fractional Fokker-Planck and diffusion-wave equations, respectively.

Eqs. \eqref{21/05/24-3} and \eqref{8/01/24-1} differ by the order of the blurred time derivative and the number of initial conditions. Eq. \eqref{21/05/24-3} needs only one initial condition
\begin{equation}\label{8/01/24-2a}
q_{s\hat{M}}({x}, t=0) = q_{0}(x),
\end{equation}
while Eq. \eqref{8/01/24-1} requires two initial conditions 
\begin{equation}\label{8/01/24-2b}
p_{s\hat{k}}(x, t=0) = p_0(x) \quad \text{and} \quad \dot{p}_{s\hat{k}}({x}, t)|_{t=0} = v_{0}(x).
\end{equation}
As the boundary conditions, we assume that solutions of Eqs. \eqref{21/05/24-3} and \eqref{8/01/24-1} vanish at $\pm$ infinity. These initial conditions mean that the diffusion process described through Eq. \eqref{21/05/24-3} results from thermal motion for $q_0(x) \geq 0$, which in the simplest case is given by the $\delta$-Dirac distribution. A similar outcome can be achieved from Eq. \eqref{8/01/24-1} when we take normalized $p_{0}(x) \geq 0$ and $v_{0}(x) = 0$ called by myself as the "diffusion-like" initial conditions. Such initial conditions were taken to solve the GFPE in Refs. \cite{FMainardi96, YuLuchko14, EBazhlekova99}. I would like to emphasize that vanishing of the second initial condition does not mean canceling its existence. Therefore, despite the non-negative and normalized form of $p_{s\hat{k}}(x, t)$, it exhibits the Doppler-like effect, see Ref. \cite{TPietrzak24, YuPovstenko22}. In general Eq. \eqref{8/01/24-1} can be solved by using $p_{0}(x) \neq 0$ and $v_{0}(x) \neq 0$ as well. Thus, the novelty of this paper relies on extending our study to the memory function beyond the power-law case and solving Eq. \eqref{8/01/24-1} under two non-zero initial conditions.

The paper is organized as follows. Based on the subordination approach in Sect. \ref{sect1} we derive the appropriate evolution operator method, with the help of which we solve the GFPE for the arbitrary smooth initial condition. In Sects. \ref{sect2} and \ref{sect3} this method is applied to solve firstly the diffusion-wave equation and, then, its generalized version. The properties of the evolution operators are studied in Sect. \ref{sect4}. The paper is concluded in Sect. \ref{sect5}. The manuscript contains seven Appendices in which the mathematical details are explained.

\section{The evolution operator method for generalized Fokker-Planck equation (GFPE)}\label{sect1}

In this section, we use the operational technique to solve the GFPE, given by Eq. \eqref{21/05/24-3}, for any form of the memory function $M(t)$. A key model illustrating how this method works is presented in Ref. \cite{KGorska12} and briefly described in the Introduction. Eq. \eqref{21/05/24-3} can be studied at least twofold, namely starting either from the subordination approach or using the Laplace-Fourier transform method.

Firstly, we begin with the subordination method \cite{KGorska23a, AVChechkin21, IMSokolov02}, {which allows one to present the solution of Eq. \eqref{21/05/24-3} in the form of
\begin{equation}\label{22/05/24-1} 
q_{s\hat{M}}(x, t) = \int_0^\infty \D\xi\, f_{s\hat{M}}(\xi, t) q_s(x, \xi), 
\end{equation} 
where 
\begin{equation}\label{22/05/24-2} 
f_{s\hat{M}}(\xi, t) = \mathscr{L}^{-1}\left[\hat{M}(s) \E^{-\xi s \hat{M}(s)}; t\right]. 
\end{equation}
The function $q_s(x, \xi)$ solves the (standard) Fokker-Planck equation and for $L_{F\!P} = B \partial_{x}^{2}$ it is equal to $N(x, \xi)$ given by Eq. \eqref{21/05/24-4}. Formally, $q_s(x, \xi)$ can be expressed as
\begin{equation}\label{22/05/24-3} 
q_s(x, \xi) = {\cal E}(1; \xi) q_0(x),  
\end{equation} 
where ${\cal E}(1; \xi)$ is introduced by Eq. \eqref{21/05/24-1} and for $L_{F\!P} = B \partial_{x}^{2}$ is equal to $\exp(t B \partial_{x}^{2})$. Inserting Eq. \eqref{22/05/24-3} into Eq. \eqref{22/05/24-1} enables one to rewrite it utilizing the evolution operator ${\cal E}_{s \hat{M}}(t)$ as follows: 
\begin{multline}\label{22/05/24-4} 
q_{s\hat{M}}(x, t) = {\cal E}_{s \hat{M}}(t) {q_0(x)}, \\ {\cal E}_{s \hat{M}}(t) = \int_0^\infty \D\xi\, f_{s \hat{M}}(\xi, t)\, {\cal E}(1; \xi), 
\end{multline} 
which for $\hat{M}(s) = 1$ yields ${\cal E}_s(t)$ equivalent to ${\cal E}(1; t)$. For $\hat{M}(s) = s^{\alpha-1}$, where $\alpha\in(0, 1)$, it simplifies to ${\cal E}(\alpha; t)$ where $f_{s\hat{M}}(\xi, t) \equiv f_{s^{\alpha}}(\xi, t)$ is equal to $f(\alpha; \xi, t)$ defined by Eq. \eqref{21/05/24-2}. (Note that the index $s \hat{M}$ informs about used memory functions $M(t)$). 

From another site, Eq. \eqref{22/05/24-4} can be derived from the Laplace and Fourier transforms of Eq. \eqref{21/05/24-3} as well. Through the Laplace and Fourier transformations, one can obtain an algebraic equation in terms of the Laplace and Fourier variables $s$ and $\kappa$. This leads to
\begin{align}\label{22/05/24-5} 
\begin{split} 
&\tilde{\hat{q}}_{{ s \hat{M}}}(\kappa, s) = \frac{\hat{M}(s) \tilde{q}_0(\kappa)}{s \hat{M}(s) + \tilde{L}_{F\!P}} \\ 
& \quad= \int_0^\infty\!\! \D\xi {\left[\hat{M}(s) \E^{-\xi s \hat{M}(s)}\right]} \left[\E^{-\xi \tilde{L}_{F\!P}} \tilde{q}_0(\kappa)\right], 
\end{split} 
\end{align} 
where $\tilde{L}_{F\!P} = \mathscr{F}^{-1}[L_{F\!P}; \kappa]$. We emphasize that this approach works for the Fokker-Planck operator whose Fourier transform can be separated from  $\tilde{q}_{s\hat{M}}(\kappa, t)$. The lower formula is derived by employing the Euler integral of the second kind, denoted as $b^{-1} = \int_0^\infty \exp(-b \xi) \D \xi$. The inverse Laplace transform applied to the first square bracket in Eq. \eqref{22/05/24-5} leads to Eq. \eqref{22/05/24-2}. The inverse Fourier transform of $\exp(-\xi \tilde{L}_{F\!P}) \tilde{q}_0(\kappa)$ yields $q_s(x, \xi)$.

In particular, for ${L}_{FP} = B \partial_{x}^{2}$ used in Eq. \eqref{22/05/24-4} with $\hat{M}(s) = s^{\alpha-1}$ (or Eq. \eqref{21/05/24-1}) we obtain
\begin{align}\label{31/05/24-1}
    & { q_{s^{\alpha}}(x, t) \equiv q(\alpha; x, t)}  \nonumber\\ & = \int_{-\infty}^{\infty} \frac{\D y}{2\sqrt{B}} \left[\int_0^\infty\!\!\! \D\xi\, {f(\alpha; \xi, t)} f\left({ \frac{1}{2};} \frac{|x-y|}{\sqrt{B}}, \xi\right)\right] q_0(y) \nonumber\\
    & = \int_{-\infty}^{\infty} \frac{\D y}{2\sqrt{B}} f\left({ \frac{\alpha}{2};} \frac{|x-y|}{\sqrt{B}}, t\right) q_0(y),
\end{align}
where { we used Eq. \eqref{21/05/24-4} whose integral kernel is rewritten as $(\pi \xi)^{-1/2} \exp[-(x-y)^{2} /(4 B \xi)] = f(1/2; |x-y|/\sqrt{B}, t)$.} The square bracket is calculated with the help of \cite[Eqs. (29)-(32)]{KAPenson16}. Obviously, for {$q_0(y) = \delta(y)$} we have 
\begin{equation}\label{31/05/24-2}
q({ \alpha;} x, t) = \frac{1}{2\sqrt{B}} f\big({\alpha/2}; |x|/\sqrt{B}, t\big), 
\end{equation}
which reconstructs the solutions of anomalous diffusion \cite{FMainardi03, FMainardi01, JKlafter12}. For ${L}_{F\!P} = B \partial_x^2 - \mu \partial_x$, Eq. \eqref{22/05/24-3} signifies: 
\begin{align}\label{22/05/24-6}
\begin{split}
& \E^{\xi (B \partial_x^2 - \mu \partial_x)} q_{0}(x) = \E^{\xi B \partial_x^2} \E^{-\xi \mu \partial_x}q_0(x) \\
& \qquad \qquad = \E^{\xi B \partial_x^2} q_0(x - \mu\xi)  = N(x - \mu\xi, \xi)
\end{split}
\end{align}
and
\begin{align}\label{28/06/24-1}
    q(\alpha; x, t)  = \int_0^\infty \D\xi {f(\alpha; \xi, t)} N(x - \mu\xi, \xi), 
\end{align}
where $N(x, t)$ is given in Eq. \eqref{21/05/24-4} and contains {$q_0(x)$}. The PDFs \eqref{28/06/24-1} with non-zero drift coefficient $\mu \neq 0$ are plotted in Fig. \ref{fig1}.
\begin{figure}
    \centering
    \includegraphics[scale=0.23]{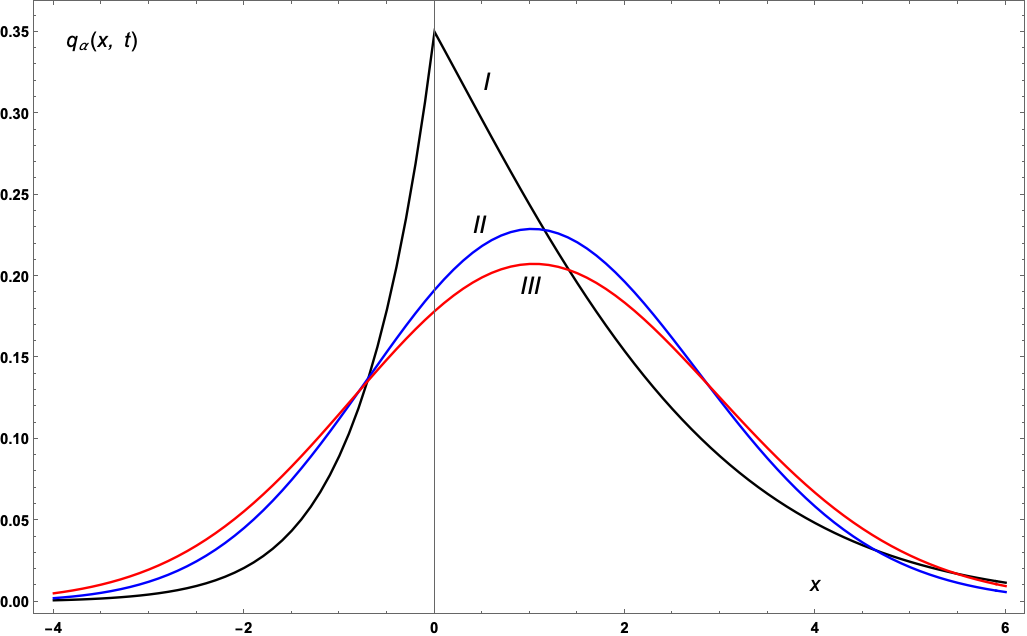}
    \caption{Plot of Eq. \eqref{28/06/24-1} for $\alpha=1/2$ and different initial conditions $q_0(x)$. The black curve no. I is for $q_0(x) = \delta(x)$, the blue curve no. II is for the box (rectangular) initial condition \eqref{13/06/24-1}, and the red curve no. III is for the Gaussian initial condition \eqref{22/01/24-1}. We take $t = 1$, $\mu = 1$, $B=1$, $\epsilon = 1$, and $\sigma_x = 1$.}
    \label{fig1}
\end{figure}

Since we begin from the subordination approach, we know that $q_s(x, \xi)$ represents a PDF of the parent process, while $f_{{ s\hat{M}}}(\xi, t)$ represents a PDF of the leading process. Moreover, for a Fokker-Planck operator containing only diffusion part $\xi$ is interpreted as an internal time. However, starting from the integral decomposition, i.e., from the inverse Laplace and Fourier transformation of the lower equation in Eq. \eqref{22/05/24-5}, we should show that $f_{s\hat{M}}(\xi, t)$ and $q_s(x, \xi)$ are normalized in the first argument and non-negative. Additionally, $\hat{f}_{{ s\hat{M}}}(\xi, s)$ should be infinitely divisible. The non-negativity and the infinite divisibility of $f_{s\hat{M}}(\xi, t)$ are ensured by the fact that the memory function $\hat{M}(s)$ is a Stieltjes function, see Lemma \ref{3}, Remarks \ref{2a} and \ref{2b} of Appendix \ref{app4}. We point out that the infinite divisibility of $\hat{f}_{s\hat{M}}(\xi, s)$ is {an essential} feature that ensures the existence of composition law for ${\cal E}_{s \hat{M}}(t)$ and it will be discussed in Sect. \ref{sect3}.  The non-negativity and normalization of $q_{s}(x, \xi)$ for $L_{F\!P}= B \partial_{x}^{2}$ and $L_{F\!P} = B \partial_{x}^{2} - \mu\partial_{x}$ flow out from Eq. \eqref{21/05/24-4} for normalized and non-negative $q_{0}(x)$. Thus, it can be called a PDF. A more general form of $q_{0}(x)$ is studied in Ref. \cite{WRSchneider89}. There is assumed that $q_{0}(x)$ is the integrable function, namely $q_{0}\in {\cal L}(\mathbb{R})$. Furthermore, in mathematical papers \cite{SDEidelman04, AAKochubei11, ANKochubei90} it is shown that the initial condition $q_{0}(x)$ can be given by a bounded or unbounded function which satisfies minimal smoothness assumptions.

\subsection{MSD calculated for $q_{s\hat{M}}(x, t)$}\label{sect1.1}

{ In this Subsection, we calculate the mean square displacement  (MSD) $(\Delta x)_{q_{s\hat{M}}} = \langle x^{2}(t) \rangle_{q_{s\hat{M}}} - \langle x(t) \rangle_{q_{s\hat{M}}}^2$.} The exact form of $q_{s\hat{M}}(x, t)$ or its Laplace and Fourier { transform}, $\tilde{\hat{q}}_{ s\hat{M}}(\kappa, s)$, enables us to compute the first $\langle x(t)\rangle_{q_{s\hat{M}}}$ (the mean (average) value) and the second $\langle x^2(t)\rangle_{q_{{ s\hat{M}}}}$  moments. { For $L_{F\!P} = B \partial_{x}^{2} - \mu \partial_{x}$} they can be calculated by taking the $m$th ($m = 1, 2$) derivative over $\kappa$ of Eq. \eqref{22/05/24-5} and, then, taking its value at $\kappa=0$. That leads to
\begin{align}\label{12/06/24-1}
\begin{split}
    \langle x(t) &\rangle_{q_{s\hat{M}}} = \mathscr{L}^{-1}\big[-\I\partial_{\kappa}\,\hat{\tilde{q}}_{\hat{M}}(\kappa, s); t\big]_{\kappa = 0} \\ 
    & = -\I \tilde{q}_0'(\kappa)\Big\vert_{\kappa=0} + \mu\,  \tilde{q}_0(0) \mathscr{L}^{-1}\big[s^{-2} \hat{M}^{-1}(s); t\big]
\end{split}
\end{align}
and
\begin{align}\label{12/06/24-2}
    \langle x^{2}(t) &\rangle_{q_{s\hat{M}}} = \mathscr{L}^{-1}\big[- \partial_{\kappa}^{2}\,\hat{\tilde{q}}_{\hat{M}}(\kappa, s); t\big]_{\kappa = 0} \nonumber\\
    & = 2 B \tilde{q}_0(0) \mathscr{L}^{-1}\big[s^{-2} \hat{M}^{-1}(s); t\big] - \tilde{q}_0{''}(\kappa)\Big\vert_{\kappa = 0}  \nonumber \\
    & + 2 \mu^2 \tilde{q}_0(0) \mathscr{L}^{-1}\big[s^{-3} \hat{M}^{-2}(s); t\big] \nonumber\\ 
    & + 2\I \mu \tilde{q}_0'(\kappa)\Big\vert_{\kappa=0}\, \mathscr{L}^{-1}\big[s^{-2} \hat{M}^{-1}(s); t\big].
\end{align}

Since the first and second moments are real, then to eliminate from them the terms with imaginary coefficient, we set $\tilde{q}'_0(\kappa)$ at $\kappa = 0$ equal to zero, i.e., we assume that $\tilde{q}_0(\kappa)$ has an extreme point at $\kappa = 0$. From the condition that the second moment is a non-negative function, it follows that $\tilde{q}_{0}(\kappa)$ at $\kappa = 0$ can be given by a non-negative function, and its second derivative at $\kappa=0 $ is negative. Moreover, taking into account the consideration presented in this section, we assume that $q_{0}(x)$ is a PDF. The initial $q_0(x)$ which has such properties are given by, e.g., $\delta$-Dirac distribution (the fundamental initial condition), the box (rectangular) function 
\begin{equation}\label{13/06/24-1}
q_{0, 1}(\epsilon; x) = \frac{1}{2\epsilon} \big[\Theta(x + \epsilon) - \Theta(x -\epsilon)\big], \quad \epsilon > 0
\end{equation}
and the Gaussian 
\begin{equation}\label{22/01/24-1}
q_{0, 2}(x) = \frac{1}{\sqrt{2\pi} \sigma_{x}} \E^{-x^{2}/(2 \sigma_{x}^{2})}, \qquad \sigma_{x} > 0.
\end{equation}

In the first case, the Fourier transform of the $\delta$-Dirac distribution is equal to 1; hence, its derivatives vanish. The Fourier transform of the box (rectangular) function reads $\tilde{q}_{0, 1}(\epsilon; \kappa) = (\sin\epsilon\kappa)/(\epsilon \kappa)$ whose limit at $\kappa\to 0$ gives 1, its first derivative at $\kappa = 0$ vanishes whereas its second derivative at $\kappa = 0$ is equal to $ -\epsilon^{2}/3$. The Fourier transform of the Gaussian gives $\tilde{q}_{0, 2}(\sigma; \kappa) = \exp(-\sigma_{x}^{2} \kappa^{2}/2)$ and at $\kappa = 0$ is equal to $1$, $\tilde{q}\,'_{0, 2}(\sigma; \kappa)|_{\kappa=0}  = 0$ as well as $\tilde{q}''_{0, 2}(\sigma; \kappa)|_{\kappa=0} = -\sigma^{2}_{x}$. Notice that for a fixed memory kernel $\hat{M}(s)$ and considering initial $q_0(x)$ the second moment $\langle x^{2}(t) \rangle_{q_{{ s\hat{M}}}}$ differs by the constant given by $\tilde{q}''_0(\kappa)|_{\kappa=0}$. We can conclude that the knowledge of the second moments does not allow one to distinguish what kind of initial conditions are used, i.e., the second moments are the same for $q_{0, 1}(x)$ and $q_{0, 2}(x)$ if we assume that $\epsilon^2/3 = \sigma_x^2$. Therefore, many forms of $q_{s \hat{M}}(x, t)$ given by Eq. \eqref{28/06/24-1} with various initial conditions lead to similar solutions. 

For instance, for the power-law $M(t)$, i.e. $M(t) = t^{-\alpha}/\Gamma(1-\alpha)$ and $\hat{M}(s) = s^{\alpha -1 }$, { the first and second moments are equal to}
\begin{equation}\label{20/06/24-2} 
\langle x(t) \rangle_{q_{s\hat{M}}} = \frac{\mu}{\Gamma(1+\alpha)} t^\alpha. 
\end{equation}
and 
\begin{equation}\label{20/06/24-1} 
\langle x^{2}(t) \rangle_{q_{s^{\alpha}}} = \frac{2 \mu^2}{\Gamma(1+2\alpha)} t^{2\alpha} + \frac{2B}{\Gamma(1+\alpha)} t^\alpha + C, 
\end{equation} 
where $\tilde{q}_0(0)=1$, $q'0(\kappa)|_{\kappa=0} = 0$, and $C = -q_0''(\kappa)|_{\kappa=0} = \text{const} \geq 0$. That implies
\begin{multline}\label{20/06/24-10}
    (\Delta x)_{q_{s^{\alpha}}} = \mu^2 t^{2\alpha} \left(\frac{2}{\Gamma(1+2\alpha)} - \frac{1}{\Gamma^2(1+\alpha)}\right) \\ + \frac{2B}{\Gamma(1+\alpha)} t^\alpha + C,
\end{multline}
where the bracket is a positive constant. For $t \ll 1$, the term proportional to $t^\alpha$ dominates, whereas for $t \gg 1$, the term containing $t^{2\alpha}$ is dominant. 

The next example is for the distributed order memory function for which $\hat{M}_d(s) = (s-1)/(s \ln s)$. From Tauberian theorem (Appendix \ref{app6}) in the limit $t\to \infty$ we find 
\begin{equation}\label{21/06/24-2}
 (\Delta x)_{q_{\hat{M}_d}} \sim \mu^2 \ln^2 t + 2 B \ln t + C. 
\end{equation}
Nevertheless, when we take $\mu = 0$ then we get
\begin{equation}\label{25/06/24-1}
    (\Delta x)_{q_{\hat{M}_d}} = \langle x^2(t) \rangle_{q_{\hat{M}_d}} \sim 2B\ln t.
\end{equation}
Based on these examples, we can suppose that in a long time limit, the terms containing $\mu^2$ control the behavior of MSD, which in this situation reads 
\begin{multline}\label{21/06/24-1}
    (\Delta x)_{q_{{ s\hat{M}}}} \sim \mu^2 \Big\{2 \mathscr{L}^{-1}\big[s^{-3} \hat{M}^{-2}(s); t\big] \\ - \Big(\mathscr{L}^{-1}\big[s^{-2} \hat{M}^{-1}(s); t\big]\Big)^2 \Big\}. 
\end{multline}

\section{Toy models: the diffusion-wave equation}\label{sect2}

Let us now examine, in a less rigorous way than in Refs. \cite{EBazhlekova19, EBazhlekova18}, the implications of using a formal analysis of the evolution operator technique for the diffusion-wave equations, which can be written in two forms. Both of these forms arise from Eq. \eqref{8/01/24-1}, where the second-order time-derivative {is convolved} with a power-law memory kernel represented either by $k_1(t) = t^{1-2\beta}/\Gamma(2-2\beta)$ ($\hat{k}_1(s) = s^{2\beta - 2}$, $\beta\in (1/2, 1]$) or $k_2(t) = t^{-\alpha}/\Gamma(1-\alpha)$ ($\hat{k}_2(s) = s^{\alpha - 1}$, $\alpha \in(0, 1]$). Consequently, we obtain the following equations:
\begin{equation}\label{12/01/24-1}
{^{C\!}D^{2\beta}_{t}} p_1(2\beta; x, t) = a^{2} \partial_{x}^{2} p_1(2\beta; x, t), \quad \beta\in(1/2, 1],
\end{equation}
called the diffusion-wave equation of the first kind (abbreviated as DWE-I), and
\begin{equation}\label{17/04/24-1}
{^{C\!} D^{\alpha + 1}_t} p_2(\alpha+1; x, t) = a^2 \partial_{x}^{2} p_2(\alpha+1; x, t), \quad \alpha\in(0, 1],
\end{equation}
named the diffusion-wave equation of the second kind (abbreviated as DWE-II). Note that Eq. \eqref{12/01/24-1} for $\beta\in(0, 1/2]$ reduces to the GFPE for the power-law memory kernel, i.e. the fractional Fokker-Planck equation. Hence, it will not be considered therein. To distinguish the range of time-derivative we introduce $\beta\in(1/2, 1]$ and $\alpha\in(0, 1]$ such that the time derivatives in Eqs. \eqref{12/01/24-1} and \eqref{17/04/24-1} are always greater than one and smaller or equal to two. In particular, when $\beta = 1/2$ and $\alpha = 0$ then Eqs. \eqref{12/01/24-1} and \eqref{17/04/24-1} reduce to the diffusion equations, respectively. While, we take $\beta = 1$ in Eq. \eqref{12/01/24-1} and $\alpha = 1$ in Eq. \eqref{17/04/24-1} then they correspond to the standard wave equations. Following \cite{EBazhlekova18} we take the ``diffusion-like"  initial conditions, i.e. $p_0(x) \geq 0$ and $v_0(x) = 0$, and as the boundary conditions we assume that $p_1(2\beta; x, t)$ and $p_2(\alpha + 1; x, t)$ vanish at $\pm$ infinity.

\subsection{Solution of DWE-I}\label{sect2.1}

In the view of \cite{KGorska12} Eq. \eqref{12/01/24-1} can be solved as
\begin{multline}\label{12/01/24-11}
p_1(2\beta; x, t) = {\cal U}(2\beta; t) p_{0}(x), \quad \text{formally} \\ {\cal U}(2\beta; t) = E_{2\beta}(a^{2} t^{2\beta}\partial_{x}^{2}).
\end{multline}
To proceed, we employ the series form of the Mittag-Leffler function $E_{2\beta}(\cdot)$ for $\beta\in(1/2, 1]$. Next, we apply its integral representation. 

Utilizing the series definition of the Mittag-Leffler function Eq. \eqref{12/01/24-11} can be written as follows:
\begin{equation}\label{16/01/24-1}
p_1(2\beta; x, t) = \sum_{r=0}^{\infty} \frac{(a t^{\beta})^{2 r}}{\Gamma(1+2\beta r)} \partial_{x}^{2r} p_{0}(x).
\end{equation}
When $p_{0}(x) = x^{n}$ ($n\in\mathbb{N}$), Eq. \eqref{16/01/24-1} yields the fractional heat polynomials introduced in Ref. \cite[Eq. (8)]{KGorska12} and denoted as ${_{2\beta}H_{n}^{(2)}}(x, a^{2} t^{2 \beta})$. They are
\begin{align}\label{16/01/24-2}
\begin{split}
    p_1(2\beta; x, t) & = {_{2\beta}H_{n}^{(2)}}(x, a^{2} t^{2 \beta}) \\
    & = n! \sum_{r=0}^{\lfloor n/2\rfloor} \frac{(a t^{\beta})^{2 r} x^{n-2r}}{\Gamma(1 + 2\beta r) (n-2r)!}.
\end{split}
\end{align}
For $\beta = 1/2$, these polynomials are known as the heat polynomials \cite{DVWidder75} which can be expressed through the Hermite $H_n$ polynomials as $(-\I a \sqrt{t})^n H_n\big(\I x/(2a\sqrt{t})\big)$. Thus, for any initial function $p_{0}(x)$ written in the power series form, $p_{0}(x) = \sum_{n\geq 0} c_{n} x^{n}$, $p_1(2\beta; x, t)$ can be expressed by using the fractional heat polynomials as follows:
\begin{equation}\label{16/01/24-3}
p_1(2\beta; x, t) = \sum_{n=0}^{\infty} c_{n}\, {_{2\beta}{H_{n}^{(2)}}}(x, a^{2} t^{2\beta}).
\end{equation}
Nevertheless, this series expansion has limited usefulness. Similarly to the series expansion applied for the conventional heat equation, it converges only for short times. As an example we consider the Gaussian initial condition given by Eq. \eqref{22/01/24-1}, where the convergence of the series in Eq. \eqref{16/01/24-3} for $\beta = 1/2$ is limited to $ 0 < t < \sigma^{2}_{x}/a^{2}$ \cite[Eq. (2)]{IMGessel05}. The solutions $p_1(2\beta; x, t)$ with well-behaved long-term behavior is obtained if for $\beta = 1/2$ we present Eq. \eqref{16/01/24-1} as $\exp(a^2 t \partial_x^2) p_0(x)$ { and use the Gauss-Weierstrass transform as is given by Eq. \eqref{21/05/24-4}.} Eq. \eqref{16/01/24-1} for $\beta = 1$ formally reads Eq. \eqref{12/01/24-7} in which $v_{0}(x) = 0$. 

Now, we are interested in finding a transformation of $E_{2\beta}(\cdot)$, $\beta\in(1/2, 1]$, that will allow us to determine $p_1(2\beta; x,t)$ for $t\in\mathbb{R}_+$. Firstly, we will achieve the integral representation of the Mittag-Leffler function $E_{2\beta}(x)$ for $\beta\in(1/2, 1]$. We use therefore \cite[Eq. (3.10)]{HJHaubold11} or \cite[Ex. 3.2]{MainardiBook} and  the known integral form of the Mittag-Leffler function for parameter $\beta\in (0, 1]$ given by, e.g., \cite{KGorska12, HPollard48, KWeron96}, namely
\begin{multline}\label{12/01/24-12}
E_{2\beta}(b^2 t^{2\beta}) = \frac{1}{2}\big[E_{\beta}(b t^\beta) + E_{\beta}(-b t^\beta)\big] \quad \text{and} \\ E_{\beta}(b t^{\beta}) = \int_0^\infty \D\xi\, f(\beta; \xi, t) \E^{b\xi},
\end{multline}
where $f(\beta; \xi, t)$ is given by Eq. \eqref{21/05/24-2}. Since $f(\beta; \xi, t)$ for $\beta = 1$ is equal to $\delta(t-\xi)$ then $E_1(bt) = \exp(bt)$ and Eq. \eqref{12/01/24-12} yields $E_2(b^2 t^2) = \cosh(b t)$. By analogy, $E_{2\beta}(b^2 t^{2\beta})$ is called $\cosh_{\beta}(b t^{\beta})$ whose series form reads \cite{ZTomovski22}
\begin{equation*}
E_{2\beta}(b^2 t^{2\beta}) = \cosh_{\beta}(b t^{\beta}) = \sum_{r=0}^{\infty} \frac{(bt^{\beta})^{2r}}{\Gamma(1 + 2\beta r)},
\end{equation*}
whereas the integral representation figures out 
\begin{equation}\label{18/01/24-1}
E_{2\beta}(b^2 t^{2\beta}) = \int_{0}^{\infty} \D \xi\, f({ \beta;} \xi, t) \cosh(b\xi).
\end{equation}
Identifying $b^{2}$ as $a^{2} \partial_{x}^{2}$ in Eq. \eqref{18/01/24-1}  we can formally represent $p_1(2\beta; x, t)$ as the action of the operator ${\cal U}(2\beta; x, t)$ given by Eq. \eqref{30/10/24-1} on $p_{0}(x)$. It contains the solution of wave equation presented by Eq. \eqref{12/01/24-7} for $v_{0}(x) = 0$, namely ${\cal U}(2; \xi) p_{0}(x)$, and allows one to state that 
\begin{equation}\label{6/11/24-1}
p_{1}(2\beta; x, t) = {\cal U}(2\beta; x, t) p_{0}(x)
\end{equation}
represents the integral decomposition containing $W(x, \xi)$ of Eq. \eqref{12/01/24-7} and $f(\beta; \xi, t)$. It reduces to the subordination approach for $p_{0}(x)$ being a PDF and $v_{0}(x) = 0$, in which $f(\beta; \xi, t)$ is interpreted as PDF and $\xi$ is called the internal time. 
\subsection{Solution of DWE-II}\label{sect2.2}

Eqs. \eqref{12/01/24-1} and \eqref{17/04/24-1} for $\beta = (\alpha + 1)/2$ pass into each other and the results obtained for DWE-I can be translated for DWE-II. Furthermore, if they are equipped with the same initial and boundary conditions, then $p_1(2\beta; x, t) = p_2(\alpha+1; x, t)$. For the ``diffusion-like'' initial and boundary conditions, we have
\begin{multline}\label{3/05/24-1}
p_2(\alpha + 1; x, t) = {\cal U}({ \alpha + 1;} t) p_{0}(x), \\ {\cal U}(\alpha + 1; t) = \int_{0}^{\infty} \D\xi\, f\left({ \frac{\alpha + 1}{2};} \xi, t\right)\, {\cal U}(1; \xi) p_{0}(x).
\end{multline}
Similarly, by Sect. \ref{sect2.1}, Eq. \eqref{3/05/24-1} for the ``diffusion-like'' initial conditions express the subordination relation in which $f(\frac{\alpha + 1}{2}; \xi, t)$ given by Eq. \eqref{21/05/24-2} subordinates the non-negative {and normalized} solution of the wave equation.

\section{Solution of GDWE via the evolution operator}\label{sect3}

In this section, we generalize the results obtained in Sect. \ref{sect2} for memory functions $k(t)$ different than power-law and arbitrary real initial conditions $p_{0}(x)$ and $v_{0}(x)$ from which, following \cite{WRSchneider89}, we assume that $p_{0}(x)$ and $v_{0}(x)$ belong to $L^1(\mathbb{R})$. To do this, we use the Laplace and Fourier transforms of Eq. \eqref{8/01/24-1}, which give an algebraic equation in terms of the Laplace $s$ and Fourier $\kappa$ coordinates: 
\begin{equation}\label{8/01/24-3} 
\hat{\tilde{p}}_{s\hat{k}}({\kappa}, s) = \frac{s \hat{k}(s)\, \tilde{p}_{0}(\kappa)}{s^{2} \hat{k}(s) + a^{2} { \kappa^{2}}}   + \frac{\hat{k}(s)\, \tilde{v}_{0}(\kappa)}{s^{2} \hat{k}(s) + a^{2} {\kappa^{2}}}. 
\end{equation} 
Then, we invert Eq. \eqref{8/01/24-3} by first calculating its inverse Laplace transform. That implies
\begin{equation}\label{5/11/24-1}
\tilde{p}_{s\hat{k}}(\kappa, t) = \tilde{g}_{s \hat{k}}(\kappa, t) \tilde{p}_{0}(\kappa) + \tilde{g}_{\hat{k}, s \hat{k}}(\kappa, t) \tilde{v}_{0}(\kappa),
\end{equation}
where
\begin{multline}\label{5/11/25-2}
\tilde{g}_{s \hat{k}}(\kappa, t) = \mathscr{L}^{-1}\left[\frac{s \hat{k}(s)}{s^{2} \hat{k}(s) + a^{2} { \kappa^{2}}}; t\right] \quad \text{and} \\
\tilde{g}_{\hat{k}, s \hat{k}}(\kappa, t) = \mathscr{L}^{-1}\left[\frac{\hat{k}(s)}{s^{2} \hat{k}(s) + a^{2} { \kappa^{2}}}; t\right].
\end{multline}
Moreover, from the property of the Laplace transform and assuming that $\tilde{g}_{s \hat{k}}(\kappa, 0) = 0$, it occurs 
\begin{equation}\label{5/11/24-3}
\tilde{g}_{s \hat{k}}(\kappa, t) = \frac{\D}{\D t} \tilde{g}_{\hat{k}, s \hat{k}}(\kappa, t),
\end{equation}
similar to the relation between Green functions used to solve the wave equation \cite[Chapter 7.8]{FWByron-v2}. The property \eqref{5/11/24-3} naturally appears for $\hat{k}(s) = 1$ for which $\tilde{g}_{s}(\kappa, t) = \cos(a t \kappa)$ and $\tilde{g}_{1, s}(\kappa, t) = (a\kappa)^{-1} \sin(a t \kappa)$. Eq. \eqref{5/11/24-3} is also satisfied for $k=s^{2\beta-2}$ for which from Eqs. \eqref{5/11/25-2} we get $\tilde{g}_{s^{2\beta-1}}(\kappa, t) = E_{2\beta}(-a^{2} \kappa^{2} t^{2\beta})$ and $\tilde{g}_{s^{2\beta-2}, \, s^{2\beta-1}}(\kappa, t) = t E_{2\beta, 2}(-a^{2} \kappa^{2}t^{2\beta})$. Then, applying the derivation of Mittag-Leffler function given by, e.g., \cite[Eq. (4.3.6)]{MainardiBook}, Eq. \eqref{5/11/24-3} is confirmed. The integral representation of $\tilde{g}_{s^{2\beta-1}}(\kappa, t)$ is presented by Eq. \eqref{18/01/24-1}. To find the integral form of $\tilde{g}_{s^{2\beta-2}, \, s^{2\beta-1}}(\kappa, t)$ we rewrite $t E_{2\beta, 2}(-a^{2} \kappa^{2}t^{2\beta})$ as 
\begin{multline}\label{5/11/24-4}
t E_{2\beta, 2}(-a^{2} \kappa^{2}t^{2\beta}) \\ = \frac{t^{1-\beta}}{2\! \I\! a \kappa}\big[E_{\beta, 2-\beta}(\I\! a \kappa t^{\beta}) - E_{\beta, 2-\beta}(-\I\! a \kappa t^{\beta})\big],
\end{multline}
where
\begin{equation*}
t^{1-\beta} E_{\beta, 2-\beta}(\I\! a \kappa t^{\beta}) = \int_{0}^{\infty} \D\xi\, e^{\I\! a \kappa\xi} \mathscr{L}^{-1}\big[s^{2\beta - 2} \E^{-\xi s^{\beta}}; t\big]
\end{equation*}
Substituting it into Eq. \eqref{5/11/24-4} we get 
\begin{align}\label{5/11/24-6}
\begin{split}
& \tilde{g}_{s^{2\beta-2}, \, s^{2\beta-1}}(\kappa, t) = t E_{2\beta, 2}(-a^{2} \kappa^{2}t^{2\beta}) \\
& \qquad = \int_{0}^{\infty} \D\xi\, \mathscr{L}^{-1}\big[s^{2\beta - 2} \E^{-\xi s^{\beta}}; t\big] \frac{\sin(a \xi \kappa)}{a\kappa}.
\end{split}
\end{align}

In the general case, to calculate Eqs. \eqref{5/11/25-2} we use Eq. \eqref{30/03-4} twice. For the term containing $\tilde{p}_0(\kappa)$ we set $\hat{G}_{1}(s) = \hat{k}^{1/2}(s)$ and $\hat{G}_{2}(s) = s \hat{k}^{1/2}(s)$ whereas for term with $\hat{v}_0(\kappa)$ we set $\hat{G}_{1}(s) = \hat{k}(s)$ and the same as in the previous case $\hat{G}_{2}(s)$. That enables us to separate the terms containing the $s$ coordinate from those that depend on the $\kappa$ coordinate. The inverse Laplace transform of Eq. \eqref{8/01/24-3} can, therefore, be expressed as
\begin{align}\label{8/01/24-4}
\begin{split}
    \tilde{p}_{s\hat{k}}({\kappa}, t) & = \int_0^\infty \D\xi\, F_{s \hat{k}^{1/2}}(\xi, t) \cos(a\xi {\kappa}) \tilde{p}_{0}(\kappa) \\ 
    & + \int_0^\infty \D\xi\, F_{\hat{k}, s\hat{k}^{1/2}}(\xi, t)\, {\frac{\sin(a\xi {\kappa})}{a\kappa}}\, \tilde{v}_{0}(\kappa),
\end{split}
\end{align}
in which $F_{s \hat{k}^{1/2}}(\xi, t)$ and $F_{\hat{k}, s\hat{k}^{1/2}}(\xi, t)$, written in the Laplace space, read 
\begin{multline}\label{8/01/24-5} 
\widehat{F}_{s \hat{k}^{1/2}}(\xi, s) = \hat{k}^{1/2}(s) \E^{-\xi s \hat{k}^{1/2}(s)} \quad \text{and} \\
\widehat{F}_{\hat{k}, s\hat{k}^{1/2}}(\xi, s) = \hat{k}(s) \E^{-\xi s \hat{k}^{1/2}(s)}.
\end{multline}
Note that for $\hat{k}(s) = s^{2\beta-2}$ they reconstruct $f(\beta; \xi, t)$ and the inverse Laplace transform in Eq. \eqref{5/11/24-6}. Furthermore, Eq. \eqref{8/01/24-4} for $p_{0}(x) = \delta(x)$ and $v_{0}(x) = 0$ gives the commonly known form of $p_{s\hat{k}}(x, t) = F_{s\hat{k}}(|x|/a, t)/(2a)$, which can be found in, e.g., \cite{TSandev19, FMainardi01}.  The function $p_{s\hat{k}}(x, t) = \mathscr{F}^{-1}[\tilde{p}_{s\hat{k}}(\kappa, t); x]$ is derived from the inverse Fourier transform of $\cos(a\xi\kappa) \tilde{p}_0(\kappa)$ and $(a\kappa)^{-1} \sin(a\xi\kappa) \tilde{v}_0(\kappa)$, which after comparing with Eq. \eqref{12/01/24-7}, yields 
\begin{align}\label{12/01/24-2} 
\mathscr{F}^{-1}\big[&\cos(a\xi\kappa) \tilde{p}_0(\kappa); x]  = \frac{1}{2}\big[p_0(x+a\xi) + p_0(x-a\xi)\big] \nonumber \\ 
& = {\cos(a \xi \partial_{x})} p_0(x) 
\end{align} 
and
\begin{align}\label{30/06/24-1}
    \mathscr{F}^{-1}&\left[{\frac{\sin(a\xi\kappa)}{a\kappa}} \tilde{v}_0(\kappa); x\right]  = \frac{1}{2a} \int_{x-a\xi}^{x+a\xi} \D y\, v_{0}(y) \nonumber\\
    & = {\sin(a \xi \partial_{x})} v_0(x).
\end{align}
The simplest example of Eq. \eqref{8/01/24-4} is for $k(t) = \delta(t)$ ($\hat{k}(s) = 1$) for which $F_s(\xi, t)$ and $F_{1, s}(\xi, t)$ reduce to the $\delta$-Dirac distribution. In this case, Eq. \eqref{8/01/24-4} is the solution of the wave equation; this is Eq. \eqref{12/01/24-7} with $p_{0}(x) \neq 0$ and $v_{0}(x) \neq 0$.

Inserting Eqs. \eqref{12/01/24-2} and \eqref{30/06/24-1} into Eq. \eqref{8/01/24-4} allows us to present $p_{s\hat{k}}(x, t)$ through the evolution operators. The resulting formula reads   
\begin{equation}\label{12/01/24-4}
p_{s\hat{k}}(x, t) = {\cal U}_{s\hat{k}}(t) p_{0}(x) + {\cal U}_{\hat{k}, s \hat{k}}(t) v_{0}(x)
\end{equation}
where the evolution operators ${\cal U}_{s\hat{k}}(t)$ and ${\cal U}_{\hat{k}, s\hat{k}}(t)$ have the following integral forms
\begin{multline}\label{8/01/24-9}
{\cal U}_{s\hat{k}}(t) = \int_{0}^{\infty}\!\!\D\xi\, F_{s \hat{k}}(\xi, t)\, {\cal U}_s(\xi) \quad \text{and} \\ {\cal U}_{\hat{k}, s\hat{k}}(t) = \int_{0}^{\infty}\!\!\D\xi\, F_{\hat{k}, s \hat{k}^{1/2}}(\xi, t)\, {\cal U}_{1, s}(\xi),
\end{multline}
where ${\cal U}_{s}(\xi) \equiv {\cal U}(2; \xi)$ is defined by Eq. \eqref{30/10/24-1} and
\begin{equation}\label{30/06/24-3}
{\cal U}_{1, s}(t) = {\sin(a\xi\partial_x)}.
\end{equation}
Note that similarly as for GFPE, the subscript $s \hat{k}$ and $\hat{k}$ inform about used the memory kernel $k(t)$. The diffusion case is obtained for the ``diffusion-like'' initial condition and $\hat{k}(s) = s^{-1}$ for which we have $F_{s^{1/2}}(\xi, t) = \mathscr{L}^{-1}[s^{-1/2} \exp(-\xi s^{1/2}); t] = \exp[-\xi^2/(4 t)]/\sqrt{\pi t}$ and, formally, we get
\begin{equation}\label{15/05/24-1}
{\cal U}_1(t) = \exp(-t a^2 \partial_x^2). 
\end{equation}

\subsection{Discussion of initial conditions}\label{sect3.1}

\noindent
{\em Subordination and integral decomposition.} \label{sect3.1.1} 

Identically to the separation of terms depending on the coordinates $t$ and $\kappa$ in $\tilde{p}_{s\hat{k}}(\kappa, t)$ its inverse Fourier transform, $p_{s\hat{k}}(x, t)$, can also be decomposed into the integral of the product of two functions one of which contains $t$ and the other $x$. Moreover, for the ``diffusion-like'' initial conditions, the lower part of Eq. \eqref{8/01/24-4} related to $v_0(x)$ disappears such that Eq. \eqref{8/01/24-4} contains only the upper part and, hereby, simplifies to the integral decomposition. Additionally, if we assume that $\hat{k}^{1/2}(s)$ is a Stieltjes function then $F_{s \hat{k}^{1/2}}(\xi, t)$ subordinates $\mathscr{F}^{-1}[\cos(a\xi\kappa) \tilde{p}_0(\kappa); x]$ given by Eq. \eqref{12/01/24-2}, see Lemma \ref{lem-6/11/24-1} in Appendix \ref{app7}. Due to \cite{TSandev19}, this assumption is satisfied for power-law and distributed order memory functions.

In the case of $v_{0}(x) \neq 0$, the second term containing $F_{\hat{k}, s\hat{k}^{12}}(\xi, t)$ appears in $p_{s\hat{k}}(\xi, t)$. Although, $F_{\hat{k}, s\hat{k}^{1/2}}(\xi, t)$ is a non-negative function it is not normalized { in the first argument, see Lemma \ref{lem-6/11/24-2} and Eq. \eqref{19/11/24-1}. To keep the normalization of $p_{s\hat{k}}(x, t)$, when $p_{0}(x)$ is a PDF, we can proceed twofold. Namely, either we assume that $v_{0}(x) = 0$ or $v_{0}(x) = -p_{0}'(x)$. The first situation corresponds to the so-called "diffusion-like" initial conditions. The latter one applied to the box (rectangular) function \eqref{13/06/24-1} and the Gaussian function \eqref{22/01/24-1} has the form
\begin{equation}\label{20/11/24-1}
v_{0, 1}(x) = \frac{1}{2\epsilon} \big[\delta(x-\epsilon) - \delta(x+\epsilon)\big] 
\end{equation}
and
\begin{equation}\label{20/11/24-2}
v_{0, 2}(x) = \frac{x}{\sqrt{2\pi} \sigma_{x}^{3}} \E^{-\frac{x^{2}}{2\sigma_{x}^{2}}}.
\end{equation}}

\medskip
\noindent
{\em MSD calculated for $p_{s\hat{k}}(x, t)$.}\label{sect3.1.2}

The additional requirement on $p_{0}(x)$ and $v_{0}(x)$ we obtain if we calculate MSDs for the solution of the GDWE. As shown in Sect. \ref{sect1.1}, the first and the second moments in the integral transform language read 
 \begin{equation}\label{18/01/24-6}
\langle x^{m}(t) \rangle _{p_{s\hat{k}}} = \mathscr{L}^{-1}\big[(-\I\! \partial_{\kappa})^{m}\,\hat{\tilde{p}}_{s\hat{k}}(\kappa, s); t\big]_{\kappa = 0},
\end{equation}
which for the mean value  ($m = 1$) gives
\begin{equation}\label{18/01/24-7}
\langle x(t) \rangle _{p_{s\hat{k}}} = -\I\! \tilde{p}'_{0}(\kappa)\Big|_{\kappa = 0} -\I\! t  \tilde{v}'_{0}(\kappa)\Big|_{\kappa = 0} 
\end{equation}
and for the second moment (MSD; $m = 2$) leads to
\begin{align}\label{18/01/24-8}
\langle & x^{2}(t) \rangle _{p_{s\hat{k}}} = 2 a^{2} \tilde{p}_{0}(0) \mathscr{L}^{-1}[s^{-3} \hat{k}^{-1}(s); t] - \tilde{p}''_{0}(\kappa)\Big|_{\kappa = 0} \nonumber\\
& + 2 a^{2} \tilde{v}_{0}(0) \mathscr{L}^{-1}[s^{-4} \hat{k}^{-1}(s); t] - t\,\tilde{v}''_{0}(\kappa)\Big|_{\kappa
 = 0}.
\end{align}

Assuming that the mean value $\langle x(t)\rangle_{p_{s\hat{k}}}$ is real, we should consider two situations. In the simplest one, which corresponds to the ``diffusion-like" initial conditions, the mean value equals zero. Here, we do not have the term related to $\tilde{v}'_{0}(\kappa)|_{\kappa=0}$ and, similar as for the solution of GFPE, we accept that $\kappa = 0$ is an extreme point of $\tilde{p}_{0}(\kappa)$. For this situation, MSD $(\Delta x)_{p_{s\hat{k}}}$ is equal to $\langle x^{2}(t) \rangle _{p_{s\hat{k}}}$. By requiring that Eq. \eqref{18/01/24-8} yields a non-negative MSD, we choose such initial $p_{0}(x)$ that the value of its Fourier transform at zero is non-negative, but its second derivative at $\kappa = 0$ is negative. The examples of functions possessing these properties are listed in Sect. \ref{sect1.1}.  For example, for the power-law memory kernel $k(t)\sim t^{1-2\beta}$, $\beta\in(1/2, 1]$ and $p_{0}(\kappa)$ given by Eqs. \eqref{13/06/24-1}, as well as \eqref{22/01/24-1}, the MSD reads
\begin{equation}\label{20/06/24-3}
    (\Delta x)_{p_{s^{2\beta}}} =   \frac{2 a^2}{\Gamma(1+2\beta)}t^{2 \beta} + C,
\end{equation}
where $C$ is defined below Eq. \eqref{20/06/24-1}. For $\hat{k}(s) = (s-1)^2/(s \ln s)^2$ the MSD in the long time limit reads \begin{equation}\label{21/06/24-3}
    (\Delta x)_{p_{(s-1)^2/[s (\ln s)^2]}} \sim 2 a^2 \ln^2 t,
\end{equation}
whereas for $\hat{k}(s) = (s-1)/(s\ln s)$ we have
\begin{equation}\label{25/06/24-2}
    (\Delta x)_{p_{(s-1)/\ln s}} \sim 2 a^2 t \ln t.
\end{equation}
(Eqs. \eqref{21/06/24-3} and \eqref{25/06/24-2} were calculated using Tauberian theorem presented in Appendix \ref{app6}.) Concluding, we can observe that for $t \gg 1$ we cannot distinguish MSD given either by Eqs. \eqref{20/06/24-10} and \eqref{20/06/24-3} or Eqs. \eqref{21/06/24-2} and \eqref{21/06/24-3}. Hence, without knowledge of the mean value $\langle x(t) \rangle$, the evolution of the measured system can be described either by the fractional Fokker-Planck equation with drift $\mu$ or by the diffusion-wave equation.

The second initial conditions $v_{0, j}(x)$, $j=1, 2$, defined by Eqs. \eqref{20/11/24-1} and \eqref{20/11/24-2} above in the Fourier space give
\begin{equation}\label{20/11/24-5}
\tilde{v}_{0, 1}(\kappa) = \I\! \frac{\sin \epsilon\kappa}{\epsilon} \quad \text{and} \quad \tilde{v}_{0,2}(\kappa) = \I\! \kappa \E^{-\kappa^2 \sigma_x^2/2},
\end{equation}
which vanish at $\kappa = 0$. At this value of $\kappa$, the first derivatives of $\tilde{v}_{0, j}(\kappa)$ are complex and equal to the minus imaginary unit, while their second derivatives also vanish. Since the mean value is not dependent on a memory function, it is proportional to time $t$ in both cases above.

\medskip
\noindent
{\em Figures.}\label{sect3.1.3} 

The importance of the initial conditions is also visible in Figs. \ref{fig2}, \ref{fig2a}, and \ref{fig3}. In Fig. \ref{fig2}, $p_{s^{2\beta}}(x, t) = p_1(2\beta; x, t)$ is illustrated for $\hat{k}(s) = s^{2\beta - 2}$, where $\beta = 3/4$, similar to DWE-I and  ``diffusion-like'' initial conditions. The auxiliary function $F_{s^{\beta}}(\xi, t) \equiv f(\beta; \xi, t)$ is defined by Eq. \eqref{21/05/24-2} in which the one argument one-sided L\'{e}vy stable distribution is given by Eq. \eqref{14/03-2a}. In Fig. \ref{fig2}, we observe a plateau, which starts to manifest two maximal points with increasing $t$, positioned symmetrically with respect to the OY axis, as can be seen in Fig. \ref{fig2a}. Thus, it can be concluded that for the ``diffusion-like'' initial conditions, there exists a certain $t$ at which the initial peak splits into two peaks that move away from each other. This observation is consistent with previous findings for $p_1(2\beta; x, t)$ with localized $p_0(x)$ as presented in, e.g., \cite{FMainardi96, YuLuchko14}. 

In Fig. \ref{fig3}, we abandoned the ``diffusion-like'' initial conditions so that $v_0(x) \neq 0$. As in the previous figures, we consider the power-law memory function for which
\begin{align}\label{5/07/24-1}
    p_{s^{2\beta}}&(x, t) = \int_0^\infty\!\!\! \D\xi\, f(\beta; \xi, t) \frac{1}{2}\big(p_0(x + at) + p_0(x - at)\big) \nonumber\\
    &+ \int_0^\infty\!\!\! \D\xi\, F_{s^{2\beta-2}, s^{\beta}}(\xi, t) \frac{1}{2a}\int_{x-a\xi}^{x+a\xi} \D y\, v_{0}(y),
\end{align}
where $F_\beta(\xi, t)$ is given by Eq. \eqref{21/05/24-2} and $F_{s^{2\beta-2}, s^{\beta}}(\xi, t) = \mathscr{L}^{-1}[s^{2\beta - 2} \exp(-\xi s^\beta); t]$ and can be expressed by the Wright-Mainardi function \cite{AApelblat21} or the Meijer $G$ function \cite{KGorska20a} implemented in the computer algebra systems like Mathematica. From \cite[Eq. (A2)]{KGorska20a} for rational $\beta = l/k\in(1/2, 1]$ it reads
\begin{equation}\label{1/07/24-1}
    F_{s^{2l/k - 2}, s^{l/k}}(\xi, t) = \frac{l^{2l/k - 2} \sqrt{kl}}{(2\pi)^{(k-l)/2}} G^{k, 0}_{l, k}\!\left(\frac{\xi^k l^l}{k^k t^l}\left\vert
    \begin{array}{c}
      \Delta(l, 2-2l/k) \\
      \Delta(k, 0)
    \end{array} \!\!  
    \right)\right.\!,
\end{equation}
where the upper and lower lists, i.e. $\Delta(l, 2-2l/k)$ and $\Delta(k, 0)$, are explained below Eq. \eqref{14/03-2a}.  
\begin{figure}
   \centering
    \includegraphics[scale=0.30]{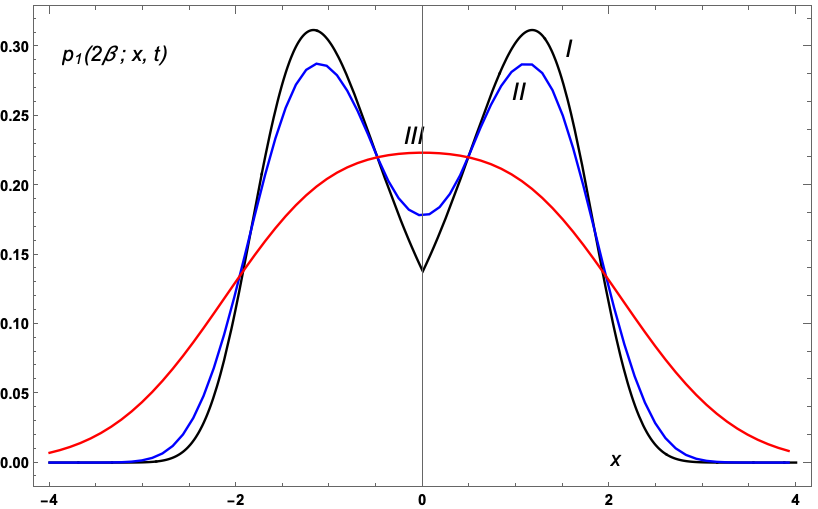}
    \caption{Plot of $p_1(2\beta; x, t)$ given by Eq. \eqref{6/11/24-1} for $\beta = 3/4$ and diffusion-like initial conditions in which $v_0(x) = 0$ and $p_0(x)$ is changing as follows: the black curve no. I is for $p_0(x) = \delta(x)$, the blue curve no. II is for the box (rectangular) function \eqref{13/06/24-1}, and the red curve no. III is for the Gaussian \eqref{22/01/24-1}. We take $t = 1$, $a=1$, $\epsilon = 1/2$, and $\sigma_x = 1$.}
    \label{fig2}
\end{figure}
\begin{figure}
    \centering
   \includegraphics[scale=0.35]{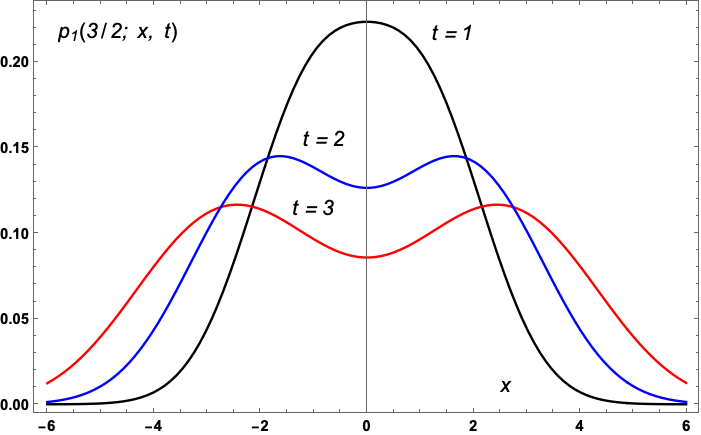}
    \caption{Plot of $p_1(2\beta; x, t)$ given by Eq. \eqref{6/11/24-1} for $\beta = 3/4$ and diffusion-like initial conditions in which $v_0(x) = 0$ and $p_0(x)$ is given by the Gaussian \eqref{22/01/24-1}. We take $a=1$, $\sigma_x = 1$, and $t = 1$ (the black curve), $t = 2$ (the blue curve), and $t = 3$ (the red curve).}
    \label{fig2a}
\end{figure}
\begin{figure}
    \centering
    \includegraphics[scale=0.26]{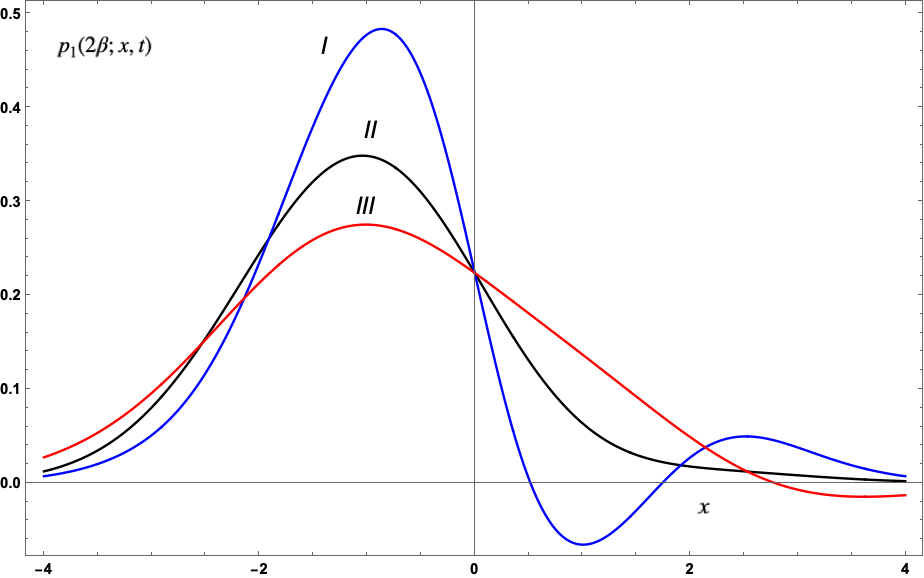}
    \caption{Plot of $p_1(2\beta; x, t)$ given by Eq. \eqref{5/07/24-1} for $\beta = 3/4$, $t = 1$ and $a = 1$. The initial $p_0(x)$ is the Gaussian \eqref{22/01/24-1} in which $\sigma_x = 1$, whereas $v_0(x)$ is given by $v_{0, 2}(x)$ given by \eqref{20/11/24-5} in which we are changing the variance $\sigma$. The blue curve no. I is for $\sigma = 0.5$, the black curve no. II is for $\sigma = 1$, and the red curve no. III is for $\sigma = 1.5$.}
    \label{fig3}
\end{figure}

\medskip
Combining results presented in Sect. \ref{sect3.1}, we get either
\begin{itemize}
\item $p_0(x)$ is given by the PDF and $v_{0}(x) = 0$, the so-called "diffusion-like" initial conditions. The mean value calculated over them is zero, so the MSD is given by the second moment $\langle x^{2}(t)\rangle_{p_{s\hat{k}}}$. The solution $p_{s\hat{k}}(x, t)$ is normalized and non-negative. These initial conditions can be used to describe anomalous diffusion; however, $p_{s\hat{k}}(x, t)$ manifests a wavelike feature, such as a frequency shift \cite{TPietrzak24}; or

\item $p_0(x)$ is given by the PDF and $v_{0}(x) = -p'_{0}(x)$. The average value calculated over these initial conditions is equal to $t$, so $(\Delta x)_{p_{s\hat{k}}} = \langle x^{2}(t)\rangle_{p_{s\hat{k}}} - t^{2}$. The choice of such initial conditions preserves the normalization of $p_{s\hat{k}}(x, t)$, but $p_{s\hat{k}}(x, t)$ may contain negative parts. Hence, they cannot be used to describe anomalous diffusion.
\end{itemize}

\section{Properties of the evolution operators: ${\cal E}_{s \hat{M}}(t)$ and ${\cal U}_{s \hat{k}}(t)$}\label{sect4}

For simplicity, in this section, we will consider only the "diffusion-like" initial conditions, which ensure that the solution of GDWE does not contain the term with $v_0(x)$. Now, the evolution operators ${\cal E}_{s \hat{M}}(t)$ and ${\cal U}_{s \hat{k}}(t)$ should satisfy the composition laws which reflect the causality properties of $q_{s\hat{M}}(x, t)$ and $p_{s\hat{k}}(x, t)$, respectively. With the help of the symbol ``$\star$'' they can be written as
\begin{equation}\label{27/05/24-1}
 {\cal E}_{s \hat{M}}(t_{2} - t_{1}) \star {\cal E}_{s \hat{M}}(t_{1} - t_{0}) = {\cal E}_{s \hat{M}}(t_{2} - t_{0}) 
\end{equation}
and
\begin{equation}\label{15/05/24-2}
 {\cal U}_{s \hat{k}}(t_{2} - t_{1}) \star {\cal U}_{s \hat{k}}(t_{1} - t_{0}) = {\cal U}_{s \hat{k}}(t_{2} - t_{0}), 
\end{equation}
where $t_0 \leq t_1 \leq t_2$. The simplest example of the composition laws \eqref{27/05/24-1} and \eqref{15/05/24-2} are for the diffusion operator, this is Eq. \eqref{22/05/24-4} for $\hat{M}(s) = 1$ and $B=1$ as well as Eq. \eqref{8/01/24-9} for $\hat{k}(s) = 1/s$ and $a=1$. Here, the symbol '$\star$' means the multiplication such that we can write it in the form
\begin{equation}\label{29/05/24-1}
 \exp\big[-(t_2 - t_1) \partial_x^2\big] \exp\big[-(t_1 - t_0) \partial_x^2\big] =  \exp\big[-(t_2 - t_0) \partial_x^2\big].
\end{equation}
The next example is the wave evolution operator ${\cal U}(2; t)$, which forms the semi-group specified in \cite[Chapter 3.14]{WArendt11}. We propose that the appropriate composition law reads
\begin{multline}\label{30/01/24-1}
\frac{\D}{\D\,(t_{2} - t_{0})} \int_{0}^{1}\!\!\! \D u \int_{t_{0}}^{t_{2}}\!\!\! \D t_{1}\, {\cal U}[2; u(t_{2} - t_{1})]\, {\cal U}[2; u(t_{1} - t_{0})] p_{0}(x) \\ = {\cal U}(2; t_{2} - t_{0}) p_{0}(x),
\end{multline}
whose derivation is presented in Appendix \ref{app5}. 

For power-law evolution operators, we distinguish two intermediate cases. One is for $\alpha \in (0, 1)$ and describes the anomalous diffusion process such that it corresponds to the operator ${\cal E}(\alpha; t)$ for $\alpha\in(0, 1]$. The next presents the evolution between the diffusion and wave behaviors. It is described by the diffusion-wave operators ${\cal U}(2\beta; t)$, $\beta \in (1/2, 1]$. Both operators, ${\cal E}(\alpha; t)$ and ${\cal U}(2\beta; t)$, are formally represented by the Mittag-Leffler functions $E_\alpha(t^\alpha L_{F\!P})$ and $E_{2\beta} (a^2 t^{2\beta} \partial_x^2)$, respectively. The use of Eq. \eqref{10/12/24-2} (Appendix \ref{app1}) enables us to write the first intermediate case describing the anomalous diffusion evolution in the form
\begin{multline}\label{1/02/24-1a}
    \frac{\D}{\D\,(t_{2} - t_{0})} \int_{0}^{1}\!\!\! \D u \int_{t_{0}}^{t_{2}}\!\!\! \D t_{1}\, {\cal E}[\alpha; u(t_{2} - t_{1})]\, {\cal E}[\alpha; u(t_{1} - t_{0})] p_0(x) \\ = {\cal E}(\alpha; t_{2} - t_{0}) p_0(x)
\end{multline}
and the next case characterizing the evolution from diffusion to wave behavior as
\begin{multline}\label{1/02/24-1b}
\frac{\D}{\D\,(t_{2} - t_{0})} \int_{0}^{1}\!\!\! \D u \int_{t_{0}}^{t_{2}}\!\!\! \D t_{1}\, {\cal U}[2\beta; u(t_{2} - t_{1})]\\ \times {\cal U}[2\beta; u(t_{1} - t_{0})] h(x) = {\cal U}(2\beta; t_{2} - t_{0}) h(x).
\end{multline}
As a starting point in the proof of Eqs. \eqref{1/02/24-1a} and \eqref{1/02/24-1b} we transform ${\cal E}(\alpha; t)$ and ${\cal U}(2\beta; t)$ into the Fourier space that gives $E_\alpha\big(t^\alpha \tilde{L}_{F\!P}(\kappa)\big)$ and $E_{2\beta}(-t^{2\beta} a^2\kappa^2)$. Thus, Eqs. \eqref{1/02/24-1a} and \eqref{1/02/24-1b} immediately flow out from Eq. \eqref{10/12/24-2}. Notice that Eq. \eqref{1/02/24-1b} for $\beta = 1$ gives Eq. \eqref{30/01/24-1}.

The fact that $f_{s \hat{M}}(\xi, t)$ and $F_{s \hat{k}^{1/2}}(\xi, t)$ are infinitely divisible functions guarantee that the evolution operators ${\cal E}_{s \hat{M}}(t)$ and ${\cal U}_{s \hat{k}}(t)$ satisfy the semi-group properties. Roughly speaking, among these semi-group properties, the most important for us is the composition law, which is related to the causality condition imposed on the solution of the generalized diffusion or diffusion-wave equations. Nevertheless, it is a challenge to find the explicit and exact form of these composition laws for memory kernels different than power laws. It is only known that the product of ${\cal E}_{s \hat{M}}(t_{2} - t_{1})$ and ${\cal E}_{s \hat{M}}(t_{1} - t_{0})$ does not give ${\cal E}_{s \hat{M}}(t_{2} - t_{0})$, namely
\begin{multline}\label{29/05/24-2}
    {\cal E}_{s \hat{M}}(t_2 - t_1) {\cal E}_{s \hat{M}}(t_1 - t_0) \\ = \int_0^\infty \D\xi \int_0^\infty \D\eta\, f_{s \hat{M}}(\xi, t_2 - t_1) f_{s \hat{M}}(\eta, t_1 - t_0) \\ \times {\cal E}(1; \xi) {\cal E}(1; \eta).
\end{multline}
From Eq. \eqref{29/05/24-1} appears that  ${\cal E}(1; \xi) {\cal E}(1; \eta) = {\cal E}(1; \xi + \eta)$. Then, setting  $\xi + \eta = v$ we have
\begin{multline}\label{29/05/24-3}
    {\cal E}_{s \hat{M}}(t_2 - t_1) {\cal E}_{s \hat{M}}(t_1 - t_0)  = \int_0^\infty \D v \int_0^v  f_{s \hat{M}}(\xi, t_2 - t_1)  \\ \times f_{s \hat{M}}(v - \xi, t_1 - t_0)  {\cal E}_1(v),
\end{multline}
compare with \cite[Eq. (18)]{KGorska12}. An { analogous} situation is for a product of the general diffusion-wave operators, i.e. ${\cal U}_{s \hat{k}}(t_{2} - t_{1})$ and ${\cal U}_{s \hat{k}}(t_{1} - t_{0})$. We point out that for this case we should use Eq. \eqref{1/02/24-2} instead of Eq. \eqref{29/05/24-1}.

However, we can express the solution of the generalized Fokker-Planck and diffusion-wave equations with the blurring represented by the memory function $M_1(t)$ and $k_1(t)$ by its solution with the other memory kernels, i.e. $M_2(t)$ and $k_2(t)$. It means that the evolution operators ${\cal E}_{s \hat{M}}(t)$ (given by Eq. \eqref{22/05/24-4}) and ${\cal U}_{s\hat{k}}(t)$ (given by Eq. \eqref{8/01/24-9}) should satisfy the self-reproducing properties:
\begin{equation}\label{2/06/24-1}
    {\cal E}_{s\hat{M}_{2} \circ s\hat{M}_{1}}(t) = \int_{0}^{\infty} \D\xi\, f_{s\hat{M}_{1}}(\xi, t)\, {\cal E}_{s \hat{M}_{2}}(\xi),
\end{equation}
and
\begin{equation}\label{23/01/24-2}
{\cal U}_{s\hat{k}_{2} \circ s\hat{k}_{1}}(t) = \int_{0}^{\infty} \D\xi\, F_{s\hat{k}_{1}}(\xi, t)\, {\cal U}_{s \hat{k}_{2}}(\xi),
\end{equation}
which flow out from Efros' theorem and are proved in Example 1 and illustrated in Example 2 of Appendix \ref{app3}. 

\section{Conclusion}\label{sect5}

Using the operational method, we solved two (1+1)-dimensional integral-differential equations of Volterra's type: GFPE and GDWE. Normalized, non-negative, and smooth functions being PDFs were selected as the initial conditions that are solutions of these equations at $t=0$, namely $q_{0}(x)$ for GFPE and $p_{0}(x)$ for GDWE. To preserve the normalization of the GDWE solution, the second initial condition, $v_{0}(x)$, is equal to either zero or the spatial derivative of $p_{0}(x)$. It was assumed that the solutions of GFPE and GDWE vanish for $x$ approaching $\pm$ infinity. The memory functions, the kernels of Volterra's equations, are denoted by $M(t)$ for the GFPE and $k(t)$ for the GDWE. 

For the considered integral-differential equations we identified two classes of evolution operators known as the generalized diffusion operator ${\cal E}_{s \hat{M}}(t)$ and the generalized diffusion-wave operators, this is ${\cal U}_{s\hat{k}}(t)$ and ${\cal U}_{\hat{k}, s \hat{k}}(t)$. {These operators are expressed by integrals containing two elements. The first of them are the integral kernels denoted as $f_{s\hat{M}}(\xi, t)$ as for GFPE or $F_{s\hat{k}^{1/2}}(\xi, t)\equiv f_{s\hat{k}^{1/2}}(\xi, t)$ and $F_{\hat{k}, s\hat{k}^{1/2}}(\xi, t)$ for GDWE. The next element is given by the operator in the case of GFPE ${\cal E}_s(\xi) \equiv {\cal E}(1; \xi)$ or in the case of GDWE ${\cal U}_s(\xi) \equiv {\cal U}(2; \xi)$ and ${\cal U}_{1,s}(\xi)$.} The functions $f_{s\hat{M}}(\xi, t)$ or $F_{s\hat{k}^{1/2}}(\xi, t)$ are normalized at the first argument, nonnegative, and their Laplace transforms are infinitely divisible. While $F_{\hat{k}, s\hat{k}^{1/2}}(\xi, t)$ preserves only the non-negativity of the above-mentioned properties.  The second element of evolution operators corresponds to the localized solutions of the considered Volterra equations. For example, for GFPE it is given by ${\cal E}(1; \xi) q_0(x) = N(x, \xi)$, where $q_0(x)$ represents the initial condition of GFPE. Thus, the solution of GFPE $q_{s \hat{M}}(x, t)$, obtained by the evolution operator method, also demonstrates the subordination approach, in which $f_{s\hat{M}}(\xi, t)$ subordinates $N(x, \xi)$. Then $q_{s \hat{M}}(x, t)$ is a PDF and can describe the diffusion process.

In the case of the GDWE equation, the situation becomes more complicated due to the presence of the second time derivative, which introduces two initial conditions, namely $p_0(x)$ and $v_0(x)$. Due to these conditions, $p_{s\hat{k}}(x, t)$ involves a two-term sum, namely ${\cal U}_{s\hat{k}}(t)p_0(x)$ and ${\cal U}_{\hat{k}, s\hat{k}}(t)v_0(x)$, which becomes identical to the terms of $W(x, t$) for $\hat{k}(s)= 1$. Hence, if $p_0(x)$ and $v_0(x)$ are non-zero, then there is no subordination approach, and negative parts of $p_{s\hat{k}}(x, t)$ appear. Even if $p_{s\hat{k}}(x, t)$ is normalized, it cannot be labeled as a PDF. Consequently, GDWE with non-zero $p_0(x)$ and $v_0(x)$ does not model diffusion phenomena. Subordination is only achieved if we include the "diffusion-like" initial conditions, i.e., set $p_0(x)$ as PDF and $v_0(x)$ as zero. 

The results presented in the paper also include the calculation of the mean values and MSDs corresponding to $q_{s\hat{M}}(x, t)$ and $p_{s\hat{k}}(x, t)$.  We take three kinds of initial condition(s) (in the case of GDWE, we use "diffusion-like" initial conditions): the $\delta$-Dirac distribution, the box (rectangular) function, and the Gaussian. On these examples we have shown that for power-law and distributed order memory functions where $\hat{M}^2(s) = \hat{k}(s)$, under "diffusion-like" initial conditions and in a long-time limit, the MSDs calculated for $q_{s\hat{M}}(x, t)$ and $p_{s\hat{k}}(x, t)$ are practically identical. Therefore, without knowing the mean values, it becomes difficult to distinguish whether our system evolves according to GPFE with drift term or according to GDWE.

We also point out that by considering the exact forms of the generalized diffusion ${\cal E}_{s \hat{M}}(t)$ and generalized diffusion-wave ${\cal U}_{s \hat{k}}(t)$ operators, we investigated their properties, particularly the composition laws that demonstrate the causal nature of the solutions of GFPE, $q_{{ s\hat{M}}}(x, t)$, and GDWE, $p_{s\hat{k}}(x, t)$. Since $\hat{f}_{\cdot}(\xi, s)$ is an infinitely divisible function, its inverse Laplace transform, $f_{\cdot}(\xi, t)$, defines the measure $\mu(\D\xi) = f_{\cdot}(\xi, t) \D\xi$, and the evolution operator expressed by it satisfies the convolution semi-group property. This agrees with the image emerging from the Green function method, in which the term containing $v_0(x)$ has been removed. When we switch on to the game $v_0(x)$ the part of the solution relevant to it involves $F_{\hat{k}, s\hat{k}^{1/2}}(\xi, t)$ which can be a non-infinitely divisible. Nevertheless, since Green function contains a propagator satisfying the semi-group properties, then $p_{s\hat{k}}(x, t)$ solving GDWE equipped with nonzero $p_0(x)$ and $v_0(x)$ should also consist of the composition of the evolution on the section $t_0 - t_1$ and $t_1 - t$, where $t_0=0 \leq t_1 \leq t$. The challenge lies in determining the explicit form of this convolution property for solutions of GPFPE and GDWE with "diffusion-like" and general initial conditions. It has been achieved for the power-law memory kernels for fractional Fokker-Planck equation and GDWE with $p_0(x) > 0$ and $v_0(x) = 0$. This leads to an open question regarding the exact and explicit form of the composition law for generalized diffusion and generalized diffusion-wave operators. 

The advantage of the method presented in the article is the possibility of choosing a spatial operator placed in the evolution operator, which for GFPE is called the Fokker-Planck operator $L_{F\!P}$. The generalized diffusion operator ${\cal E}_{s\hat{M}}(t)$ is expressed by the exponential operator  $\exp(\xi L_{F\!P})$ which act on the initial condition $q_{0}(x)$. Such action is often known; see Ref. \cite{dzielo}. For example, $L_{F\!P} = A \partial_{x}^{2} + B \partial_{x} x$, $A, B > 0$, is used in the physics of storage rings to model the effect of diffusion and attenuation of the electron beam caused by synchrotron radiation emitted by electrons in the bending magnets of {the ring \cite{FCiocci00}}. { Choosing the Fokker-Planck operator $L_{F\!P}$ in the form $A \partial_x^2 + B x^2$ we can calculate ${\cal E}(1; \xi) q_0(x)$ by using the Zassenhaus formula.} Additionally, the exponential form of the evolution operator demonstrates its semi-group properties. We believe that in the case of GDWE, we can proceed similarly; instead of the second spatial derivative, we can use a more complicated operator. However, the problem of the pseudo-operator $\sqrt{L_{F\,P}}$ and ambiguities related to it will arise.

\section*{Acknowledgment}
I express my gratitude to the anonymous Referees for carefully reading the manuscripts and for all their remarks and suggestions, which allowed me to substantially amend the paper.

I also acknowledge the financial support provided under the NCN Research Grant Preludium Bis 2 No. UMO-2020/39/O/ST2/01563.

\appendix
\section{Useful formulaes}\label{app1}

The fractional derivative in the Caputo sense reads
\begin{equation}\label{10/12/24-1}
    \big({^{C\!} D^\nu_t} f\big)(t) = \int_0^t \frac{f^{(n)}(\xi)}{(t-\xi)^{\nu + 1- n}}\, \frac{\D \xi}{\Gamma(n - \nu)}, 
\end{equation}
for $\nu\in(n-1, n)$, $n\in\mathbb{N}$.

Since the last (unnumbered) formula on page 1720 of \cite{KGorska19} is applied in Sect. \ref{sect4} we quote it below:
\begin{align}\label{10/12/24-2}
\frac{\D}{\D T_{2}} \int_{0}^{1} \D u & \int_{T_{0}}^{T_{2}} \D T_{1}\, E_{\alpha}[-u(T_{2} - T_{1})^{\alpha}]\, E_{\alpha}[-u(T_{1} - T_{0})^{\alpha}] \nonumber \\
& = E_{\alpha}[-(T_{2} - T_{0})^{\alpha}].
\end{align}

\section{One-sided L\'{e}vy stable distribution}\label{app2}

The one-sided L\'{e}vy stable distribution is defined through the inverse Laplace transform as follows
\begin{equation*}
\varPhi_{\alpha}(\sigma) = \mathscr{L}^{-1}[\E^{-s^{\alpha}}; s] = \int_{L} \E^{s \sigma}\E^{-s^{\alpha}} \frac{\D\sigma}{2\pi\!\I},
\end{equation*}
where the integral contour is described in \cite{HPollard46}. For rational $\alpha = l/k$, the one-sided L\'{e}vy stable distribution can be expressed by the Meijer G-function and/or the finite sum of hypergeometric function, see \cite[Eq. (31)]{KGorska21b}, respectively. The exact and explicit form of  $\varPhi_{l/k}(\sigma)$ reads
\begin{equation}\label{14/03-2a}
\varPhi_{l/k}(\sigma) = \frac{\sqrt{l k}}{(2\pi)^{(k-l)/2}} \frac{1}{\sigma} G^{k, 0}_{l, k}\left(\frac{l^{l}}{k^{k} \sigma^{l}}\Big| 
\begin{array}{c}
   \Delta(l, 0)\\
   \Delta(k, 0)
\end{array}
\right),
\end{equation}
where $\alpha = l/k$ such that $l < k$ and $l, k = 1, 2, \ldots$. According to common convention the special list of $n$ elements is equal to $\Delta(n, a) = a/n, (a + 1)/n, \ldots, (a + n-1)/n$ which is placed in upper (it is $\Delta(l, 0)$) and lower (it is $\Delta(k, 0)$) list of parameters. The simplest example of Eq. \eqref{14/03-2a} is given for $l/k = 1/2$ which is called the L\'{e}vy-Smirnov distribution and reads
\begin{equation*}
    \varPhi_{1/2}(\sigma) = \frac{1}{2\sqrt{\pi}\sigma^{3/2}} \exp\left(-\frac{1}{4\sigma}\right).
\end{equation*}

\section{Efros' theorem}\label{app3}

Efros' theorem \cite{AMEfros35, LWlodarski52, UGraf04, KGorska12a, AApelblat21} generalizes the convolution theorem for the Laplace transform. It states as follows:
\begin{theorem}\label{t1}
If $\hat{G}_{1}(s)$ and $\hat{G}_{2}(s)$ are analytic functions, and $\mathscr{L}[h(x, \xi); s] = \hat{h}(x, s)$ as well as $\mathscr{L}[f(\xi, t); s] = \int_{0}^{\infty} f(\xi, t) \E^{-z t} \D t = \hat{G}_{1}(s) \E^{-\xi \hat{G}_{2}(s)}$ exist, then 
\begin{equation*}
\hat{G}_{1}(s)\hat{h}(x,\hat{G}_{2}(s)) = \int_{0}^{\infty}\!\Big[\!\int_{0}^{\infty}\!\! f(\xi, t) h(x, \xi) \D\xi\Big] \E^{- s t} \D t.
\end{equation*}
\end{theorem}
\noindent
From Efros' theorem appears that 
\begin{equation}\label{30/03-4}
\mathscr{L}^{-1}[\hat{G}_{1}(s)\hat{h}(x, \hat{G}_{2}(s)); t] = \int_{0}^{\infty}\!\! f(\xi, t) h(x, \xi) \D\xi.
\end{equation}

\noindent
{\bf Example 1} 
This theorem can be exemplified by taking $\hat{M}(s)$ and the operator ${\cal E}_{s \hat{M}}(t)$ defined as
\begin{multline}\label{20/06/24-5}
    {\cal E}_{s \hat{M}}(t) = \int_0^\infty \D\xi\, f_{s\hat{M}}(\xi, t) {\cal E}(1; \xi), 
        \\ f_{s\hat{M}}(\xi, t) = \mathscr{L}^{-1}\big[\hat{M}(s) \E^{-\xi s \hat{M}(s)}; t\big].
\end{multline}
The formula
\begin{equation}\label{20/06/24-6}
     {\cal E}_{s \hat{M}_1 \circ s \hat{M}_2}(t) = \int_0^\infty \D\xi\, f_{s\hat{M}_1}(\xi, t) {\cal E}_{s \hat{M}_2}(\xi)
\end{equation}
can be proved by inserting Eq. \eqref{20/06/24-5} into it which gives
\begin{multline}\label{23/01/24-3a}
\int_{0}^{\infty} \D\xi\, f_{s\hat{M}_{1}}(\xi, t)\, \left[\int_{0}^{\infty} \D y\, f_{s\hat{M}_{2}}(y, \xi)\, {\cal E}(1; y)\right] \\
= \int_{0}^{\infty}\!\!\! \D y \left[\int_{0}^{\infty} \D\xi f_{s\hat{M}_{2}}(y, \xi) f_{s\hat{M}_{1}}(\xi, t)\right] {\cal E}(1; y).
\end{multline}
From Efros' theorem, in which $\hat{G}_1(s) = \hat{M}_{1}(s)$ and $\hat{G}_2(s) = s \hat{M}_{1}$, Eq. \eqref{30/03-4} shows that
\begin{multline*}
\int_{0}^{\infty} \D\xi \mathscr{L}^{-1}\Big[{\hat{M}_{2}(s)} \E^{\!-y s {\hat{M}_{2}(s)}}; \xi\Big] \mathscr{L}^{-1}\Big[{\hat{M}_{1}(s)} \E^{\!-\xi s {\hat{M}_{1}(s)}}; t\Big] \\ = \mathscr{L}^{-1}\Big\{\hat{M}_{1}(s) \hat{M}_{2}\big(s \hat{M}_{1}(s)\big) \E^{-y s {\hat{M}_{1}(s)}{\hat{M}_{2}\big(s {\hat{M}_{1}(s)}\big)}}; t\Big\},
\end{multline*}
which if written in a more compact form has the form
\begin{equation}\label{23/01/24-4}
\int_{0}^{\infty} \D\xi f_{s\hat{g}_{2}}(y, \xi) f_{s\hat{g}_{1}}(\xi, t) = f_{s\hat{g}_{2} \circ s\hat{g}_{1}}(y, t).
\end{equation}
Substituting it into Eq. \eqref{23/01/24-3a} and using the definition of evolution operator \eqref{20/06/24-5} we complete the proof. 

The same consideration can be made for ${\cal U}_{s \hat{k}}(t)$.

\noindent
{\bf Example 2} 
The example of Eq. \eqref{2/06/24-1} for both memory functions $M_1(t)$ and $M_2(t)$ given by the power-laws are exhibited in \cite[Eq. (17)]{KGorska12}. Here, we illustrate Eq. \eqref{2/06/24-1} when one of the memory functions $M_1(t)$ or $M_2(t)$ has a power-law form. At first, we take that $\hat{M}_1(s) = s^{\alpha-1}$ then Eq. \eqref{23/01/24-4} reads
\begin{equation}\label{13/06/24-3}
    \int_0^\infty \D\xi\, f_{s \hat{M}_2} (y, \xi) f_\alpha(\xi, t) = f_{s^\alpha \hat{M}_2(s^\alpha)}(t)
\end{equation}
and Eq. \eqref{2/06/24-1} yields
\begin{equation}\label{13/06/24-4}
    {\cal E}_{s^\alpha \hat{M}_2(s^\alpha)} = \int_0^t \D\xi\, f_\alpha(\xi, t) {\cal E}_{s \hat{M}_2}(\xi).
\end{equation}
Assuming that $\hat{M}_2(s) = s^{\alpha - 1}$ such that Eqs. \eqref{23/01/24-4} and \eqref{2/06/24-1}, respectively, figure out
\begin{multline}\label{13/06/24-5}
    \int_0^\infty \D\xi\, f(\alpha; y, \xi) f_{s \hat{M}_1}(\xi, t) = f_{(s \hat{M}_1)^\alpha}(t) \quad \text{and} \\
    {\cal E}_{(s \hat{M}_1)^\alpha}(t) = \int_0^\infty \D\xi\, f_{s \hat{M}_1}(\xi, t) {\cal E}(\alpha; \xi).
\end{multline}
Similar results can be obtained for evolution operator ${\cal U}_{s \hat{k}}$.

\section{Definitions and proofs of Sect. \ref{sect1}  statements}\label{app4}

Among non-negative $C^\infty$ functions on $\mathbb{R}_{+}$, we can distinguish the complete monotone and Bernstein functions. According to \cite{RLSchilling10}, their definitions and subclasses are given below.  

\begin{definition}\label{CMF}
{\rm
A function $\hat{c}: [0, \infty) \mapsto (0, \infty)$ is {\em completely monotone function (CMF)} if its derivatives exist and $(-1)^n \hat{c}^{(n)}(s) \geq 0$, for all $n\in\mathbb{N}_0$. }
\end{definition}
The important feature of any CMF is that Bernstein's theorem uniquely connects it with a non-negative function $c(t)$ through the Laplace integral
\begin{equation*}
    \hat{c}(s) = \int_0^\infty \E^{-st} c(t) \D t.
\end{equation*}
\begin{definition}\label{SF}
{\rm A (non-negative) {\em Stieltjes function (SF)} is a function $\hat{g}: (0, \infty) \mapsto (0, \infty)$ which can be written as
\begin{equation*}
    \hat{g}(s) = \frac{A}{s} + B + \int_0^\infty \frac{\sigma(\D t)}{s + t},
\end{equation*}
where $A, B \geq 0$ and $\sigma$ is a measure on $(0, \infty)$ such that $\int_0^\infty \sigma(\D t)/(1+t) < \infty$. }
\end{definition}
Note that the class of SFs belongs to the class of CMFs and, thus, it is a decreasing function. Usually, it is assumed $A = 0$ and $B = 0$ which simplifies the calculations and ensures its integrability at zero and vanishing at infinity.

\begin{definition}\label{BF}
{\rm 
A function $\hat{b}: [0, \infty) \mapsto (0, \infty)$ is {\em Bernstein function (BF)} if its all derivatives exist, $\hat{b}(s)$ is non-negative, $\hat{b}'(s)$ is a CMF. If a BF $\hat{b}: (0, \infty) \mapsto (0, \infty)$, and, additionally, $\hat{b}(s)/s$ is a SF then it is called a {\em completely Bernstein function (CBF)}. }
\end{definition}
\begin{definition}\label{lCMF}
{\rm
A function $\hat{l}: (0, \infty) \mapsto (0, \infty)$ is a {\em logarithmically complete monotone function (lCMF)} if $-(\ln \hat{l})'$ is a CMF. }
\end{definition}

The characteristics of these functions and their properties can be referenced in \cite{RLSchilling10}. The key features highlighted in this paper include (i) the product or sum of CMFs results in another CMF; (ii) the composition of CMF and BF(CBF) yields a CMF; (iii) for the CBF, there exists a corresponding partner $\hat{b}^{\star}$ such that $\hat{b}(s) \hat{b}^{\star}(s) = s$; and (iv) the algebraic inverse of SF (CBF) is CBF (SF).  

\begin{lemma}\label{1}
   {\rm For $\hat{g}(s)$ being a SF emerges that $s \hat{g}(s)$ is a CBF.}
\end{lemma}
\noindent
{\em Proof.} Lemma \ref{1} comes immediately from the fact that $\hat{g}(s)$ is a SF and from property (iv) according to which $1/\hat{g}(s) = s/[s \hat{g}(s)]$ is a CBF and from the property (iii). 

\begin{lemma}\label{3}
    {\rm $\hat{f}_{s \hat{g}}(\xi, s) = \hat{g}(s) \exp[-\xi s \hat{g}(s)]$, for $\hat{g}(s)$ being a SF, is lCMF.}
\end{lemma}
\noindent
{\em Proof.} Lemma \ref{3} can be shown by making the direct calculations. At first, we simplify the expression on $\hat{f}_{s \hat{g}}(\xi, s)$ as follows:
\begin{equation*} 
\ln\big[\hat{g}(s) \exp(-\xi s \hat{g}(s))\big] = - \left(\ln \frac{1}{\hat{g}(s)}\right) - \xi s \hat{g}(s). 
\end{equation*}
Then, taking its derivative over $s$ and taking care for the negative sign, we obtain:
\begin{equation*} 
- \big[\ln\big(\hat{g}(s) \E^{-\xi s \hat{g}(s)}\big)\big]' = \hat{g}(s) \left(\frac{d}{ds} \frac{1}{\hat{g}(s)}\right) + \xi \frac{d}{ds} \big(s \hat{g}(s)\big), 
\end{equation*}
$\xi \geq 0$. Since for $\hat{g}(s)$ being a SF both functions $\hat{g}^{-1}(s)$ and $s \hat{g}(s)$ are CBFs, their derivatives are CMFs. Finally, applying property (i) completes the proof. 

\begin{remark}\label{2a}
{\rm \cite[Theorem 5.11]{RLSchilling10} says that if $\hat{H}: (0, \infty) \mapsto (0, \infty)$ is a lCMF then $\hat{H}$ is infinitely divisible and CMFs. From this theorem and lemma \ref{3} appears that $\hat{f}_{s \hat{g}}(\xi, s)$ is an infinitely divisible.}
\end{remark}
\begin{remark}\label{2b}
{\rm \cite[Theorem 5.11]{RLSchilling10} ensures also that $\hat{f}_{s\hat{g}}(\xi, s)$ is a CMF and the Bernstein theory implies that $f_{s\hat{g}}(\xi, t)$ is a non-negative function in $t$. }
\end{remark}

\section{Proofs in Sect. \ref{sect3}}\label{app7}

\begin{lemma}\label{lem-6/11/24-1}
{\rm
For the ``diffusion-like'' initial conditions, i.e. $p_{0}(x) > 0$, $v_{0}(x) = 0$, and $\hat{k}^{1/2}(s)$ being a Stieltjes function (SF), $p_{s\hat{k}}(x, t)$ given by Eq. \eqref{12/01/24-4} is a subordination approach. }
\end{lemma}
\noindent
{\em Proof.}
To show that $p_{s\hat{k}}(x, t)$ can be expressed in terms of the subordination approach, we need to show that both the functions $F_{s \hat{k}^{1/2}}(\xi, t)$ and $\mathscr{F}^{-1}[\cos(a\xi\kappa) \tilde{p}_0(\kappa); x]$ given by Eq. \eqref{12/01/24-2} are normalized, non-negative, and, additionally,  $\hat{F}_{s \hat{k}^{1/2}}(\xi, s)$ is infinitely divisible. 

If $\hat{k}^{1/2}(s)$ is a SF, then $s\hat{k}^{1/2}(s)$ is a CBF and, according to Bernstein's theorem, $F_{s \hat{k}^{1/2}}(\xi, t)$ is a non-negative function. Furthermore, the function $\hat{F}_{s \hat{k}^{1/2}}(\xi, s)$ exhibits lCMF (Lemma \ref{3} of Appendix \ref{app4}) and \cite[Theorem 5.11]{RLSchilling10} implies that it is both completely monotonic and infinitely divisible, see Remarks \ref{2a} and \ref{2b} of Appendix \ref{app4}. 

For the ``diffusion-like'' initial conditions, Eq. \eqref{12/01/24-2} is a normalized and non-negative function. 

\begin{lemma}\label{lem-6/11/24-2}
{\rm For $\hat{k}(s)$ being the CMF, $F_{\hat{k}, s\hat{k}^{1/2}}(\xi, t)$ given by the inverse Laplace transform of Eq. \eqref{8/01/24-5} is a non-negative function. It is not normalized, and its Laplace transform is not lCMF.} \end{lemma}
\noindent
{\em Proof.}
Due to property (i) in Appendix \ref{app4}, $\hat{F}_{\hat{k}, s\hat{k}^{1/2}}(\xi, s)$ is a CMF and Bernstein's theorem implies the non-negativity of $F_{\hat{k}, s\hat{k}^{1/2}}(\xi, t)$. The lack of normalization in the first argument stems from Eq. \eqref{8/01/24-5}.  Its inverse Laplace transform allows one to write
\begin{equation}\label{19/11/24-1}
\int_0^\infty \D\xi\, F_{\hat{k}, s\hat{k}}(\xi, t) = \mathscr{L}^{-1}\left[\frac{\hat{k}^{1/2}(s)}{s}; t\right].
\end{equation}
Hence, we can conclude that Lemma \ref{3}  of Appendix \ref{app4} is not satisfied.

\section{Tauberian's theorem \cite{Feller}}\label{app6}

\begin{theorem}\label{th2}
{\rm
If the Laplace transform pair $\hat{r}(s)$ of the function $r(t)$ behaves for $s\ll 1$ like $\hat{r}(s) \sim s^{-\rho} L(s^{-1})$, $\rho >0$, where $L(t)$ is a slowly varying function at infinity, then $r(t)$ has the asymptotic behavior at $t \gg 1$ given by $r(t) \sim t^{\rho - 1} L(t)/\Gamma(\rho)$.}
\end{theorem}

The slow varying function at infinity means that $\lim_{t\to\infty} L(at)/L(t) = 1$ for $a>0$.

\section{Derivation of Eq. \eqref{30/01/24-1}}\label{app5}

According to \cite[Lemma 3.14.3]{WArendt11}
\begin{multline}\label{1/02/24-2}
{\cal U}[2; u(t_{2} - t_{1})]\, {\cal U}[2; u(t_{1} - t_{0})] \\= \frac{1}{2}\big\{{\cal U}[2; u(t_{2} - t_{0})] + {\cal U}[2; u(t_{2} + t_{0} - 2 t_{1})]\big\}.
\end{multline}
Then, we substitute it into Eq. \eqref{30/01/24-1} and take the Fourier transform of obtained formula. Its the left-hand-side (LHS) is expressed as
\begin{multline}\label{30/01/24-2}
\frac{1}{2}\frac{\D}{\D\,(t_{2} - t_{0})} \int_{0}^{1}\!\!\! \D u \int_{t_{0}}^{t_{2}}\!\!\! \D t_{1}\,\Big\{\cos\big[au(t_{2} - t_{0})\kappa\big]  \\ + \cos\big[au(t_{2} + t_{0} - 2 t_{1})\kappa\big]\Big\} \tilde{h}(\kappa).
\end{multline}
Calculating the appropriate integrals and derivative we get that Eq. \eqref{30/01/24-2} is equal to $\cos\big[a(t_{2} - t_{0})\kappa\big] \tilde{h}(\kappa)$ which after inverse Fourier transform yields ${\cal U}_{1}(t_{2} - t_{0}) h(x)$.


\begin{thebibliography}{99}
\bibitem{F0}
The rigorous mathematical discussion on the existence of the evolution operator ${\cal E}(\alpha; t)$ is presented in Ref. \cite{EBazhlekova19}. There is also the link between the evolution operator method and the subordination approach used for finding the solution of the fractional Fokker-Planck equation.
\end{thebibliography}
\end{document}